\documentclass[11pt]{amsart}
\usepackage{amsmath,amsfonts, amssymb, amsthm}
\usepackage{graphicx}
\usepackage{tikz}
\usepackage[margin=1.25in]{geometry}

\newtheorem{theorem}{Theorem}[section]

\numberwithin{equation}{section}
\newtheorem{lemma}[theorem]{Lemma}
\newtheorem{example}[theorem]{Example}

\newtheorem{corollary}[theorem]{Corollary}

\newcommand{\R}{{\mathbb R}}
\newcommand{\bR}{{\mathbb R}}
\newcommand{\C}{{\mathbb C}}

\newcommand{\Cx}{{\mathbb C}}
\newcommand{\E}{{\mathbb E}}

\newcommand{\bZ}{{\mathbb Z}}
\newcommand{\Ir}{{\mathbb Z}}
\newcommand{\A}{{\mathcal A}}
\newcommand{\cA}{{\mathcal A}}
\newcommand{\cB}{{\mathcal B}}
\newcommand{\cH}{{\mathcal H}}
\newcommand{\rd}{{\rm d}}
\newcommand{\cU}{{\mathcal U}}

\newcommand{\ket}[1]{\left\vert #1\right\rangle}

\renewcommand{\Im}{\,\mathrm{Im}\,}   
\def\idty{{\mathchoice {\mathrm{1\mskip-4mu l}} {\mathrm{1\mskip-4mu l}} %
{\mathrm{1\mskip-4.5mu l}} {\mathrm{1\mskip-5mu l}}}}

\newcommand{\Tr}{\mbox{Tr}}

\newcommand{\supp}{\mathop{\mathrm{supp}}}

\newcommand{\be}{\begin{equation}}
\newcommand{\ee}{\end{equation}}
\newcommand{\bea}{\begin{eqnarray}}
\newcommand{\eea}{\end{eqnarray}}
\newcommand{\beann}{\begin{eqnarray*}}
\newcommand{\eeann}{\end{eqnarray*}}
\newcommand{\eq}[1]{(\ref{#1})}

\begin{document}
\renewcommand{\thefootnote}{\fnsymbol{footnote}}
\title[Lieb-Robinson Bounds]{Lieb-Robinson Bounds in Quantum Many-Body Physics}

\author[B. Nachtergaele]{Bruno Nachtergaele}
\address{Department of Mathematics\\
University of California, Davis\\
Davis, CA 95616, USA}
\email{bxn@math.ucdavis.edu}

\author[R. Sims]{Robert Sims}
\address{Department of Mathematics\\
University of Arizona\\
Tucson, AZ 85721, USA}
\email{rsims@math.arizona.edu}

\date{Version: \today }
\bigskip
\begin{abstract}
We give an overview of recent results on Lieb-Robinson bounds and some
of their applications in the study of quantum many-body models in condensed
matter physics.
\end{abstract}

\maketitle
\footnotetext[1]{Copyright \copyright\ 2010 by the authors. This
paper may be reproduced, in its entirety, for non-commercial
purposes.}

\section{Introduction}\label{sec:intro}

Condensed matter physics and quantum computation have a common interest in
`complicated' states of quantum many-body systems. Low-temperature physics 
is often described by states with strong correlations and many interesting
physical properties are due to those correlations. Quite often, these correlations
are intimately related to fundamental features of quantum mechanics in a phenomenon
that is usually referred to as {\em entanglement}. In the case of quantum computation
entanglement is the essential feature without which the subject would not exist.

The aim of these lecture notes is to present recent results that emerged from
this shared interest in entangled quantum states. Entanglement is often described
as `non-local correlations' that are of a fundamentally different nature than
correlations in the classical sense. The latter can be described as a property of 
a multivariate probability distribution. While it is absolutely correct that quantum 
entanglement is something new that is not found in any classical description of
physics, it would not be correct to say that the complexity implied by entanglement 
precludes a useful discussion of locality in quantum physics.  One of the key tools to 
unravel the complexity of typical ground states of a condensed matter physics model,
states that could potentially be of interest for the implementation of a quantum
computer, are the so-called {\it Lieb-Robinson bounds}. Lieb-Robinson bounds
imply that non-relativistic quantum dynamics has, at least approximately, the same
kind of locality structure provided in a field theory by the finiteness of the speed of
light. 

The original work by Lieb and Robinson dates back to 1972 \cite{lieb:1972}.
Since the work of Hastings \cite{hastings:2004} there have been a series of extensions 
and improvements \cite{nachtergaele:2006a, hastings:2006,nachtergaele:2006,
nachtergaele:2009a,nachtergaele:2009b, raz:2009}. These extension were largely
motivated by the possibility of applying Lieb-Robinson bounds to a variety of problems
concerning correlated and entangled states of quantum many-body systems. Other applications 
are focused on the dynamics of such systems. For example, a Lieb-Robinson type estimate
was used in \cite{erdos:2009} to derive the nonlinear Hartree equation from mean-field
many body dynamics. In \cite{nachtergaele:2010}, the application is to prove the existence of
the dynamics of a class of anharmonic oscillator lattices in the thermodynamic limit.

In these lecture notes we will discuss Lieb-Robinson bounds, also called 
{\it locality} or {\it propagation estimates}, depending on one's point of view,
for quantum lattice systems and some of the applications in which they play a central role.
Section \ref{sec:lr} is devoted to the Lieb-Robinson bounds themselves. We start
by considering systems with bounded interactions, of which quantum spin systems
are the typical examples, but which also includes models with an infinite-dimensional
Hilbert space on each site, as long as the only unbounded terms in the Hamiltonian
act on single sites. An example of this situation is the quantum rotor model.
Next, in Section \ref{sec:unbounded} we study a class of systems with unbounded
interactions: the harmonic oscillator lattice models and suitable anharmonic perturbations
of them.

In Section \ref{sec:existence} we show how Lieb-Robinson bounds can be used to 
prove the existence of the dynamics in the thermodynamic limit. Intuitively, it is easy to
understand that a locality property of some kind is essential for the dynamics of
an infinite system of interacting particles to be well-defined. Again, we first treat systems
with bounded interactions only, and then turn to the anharmonic oscillator lattices systems
for which new subtleties arise.

The first direct application to ground states of quantum many-body systems
is the exponential clustering theorem which, succinctly stated, says that a
non-vanishing spectral gap above the ground state implies the existence 
of a finite correlation length in that ground state. We sketch a proof of this
results in Section \ref{sec:gapped_gs}. One can obtain much more detailed
information about the structure of the ground state. A good example of this is
the Area Law for the entanglement entropy, which Hastings proved for 
one-dimensional systems \cite{hastings:2007}. We have made some progress to
understand the structure of gapped ground states in higher dimensions, but 
a general Area Law is still lacking. 
The structure of gapped ground states that
is now revealing itself can be regarded as a generalization of a special type
of states that were shown to be the exact ground states of particular Hamiltonians,
called valence bond solid models, of which the first non-trivial examples were
introduced by Affleck, Kennedy, Lieb, and Tasaki in 1987 \cite{affleck:1987a},
most notable the spin-1 chain now called the AKLT model.
These examples were generalized considerably in \cite{fannes:1992}, where
the class of finitely correlated states was introduced, a special subclass of 
which was later dubbed matrix product states. Since it is now becoming clear
that the special structure of these states is approximately present in all gapped
ground states, we provide in this section also a brief overview of the AKLT
ground state.

New applications and extensions of Lieb-Robinson bounds and techniques that
employ them continue to be found. In Section \ref{sec:conclusion}, we
briefly mention a few examples of recent results that are not discussed 
in detail in these notes.

\section{Lieb-Robinson Bounds}\label{sec:lr}

\subsection{Bounded Interactions} \label{subsec:bdint}

In this section, we will consider general quantum systems defined on a finite sets 
$\Gamma$ equipped with a metric $d$. We introduce these systems as follows.

To each site $x \in \Gamma$, we will associate a Hilbert space $\mathcal{H}_x$.
In the context of quantum spins systems, the Hilbert space $\mathcal{H}_x$ is
finite dimensional, whereas for oscillator systems, typically $\mathcal{H}_x = L^2( \R, \rd q_x)$.
For the locality results we present in this section, both systems can be treated within the
same framework.

With any subset $X \subset \Gamma$,
the Hilbert space of states over $X$ is given by
\be \label{eq:hilX}
\mathcal{H}_{X} \, = \, \bigotimes_{x \in X} \mathcal{H}_x,
\ee
and the algebra of local observables over $X$ is then defined to be
\be \label{eq:algX}
\mathcal{A}_{X} = \bigotimes_{x \in X} \cB (\cH_x),
\ee
where $\cB (\cH_x)$ denotes the algebra of bounded linear operators on $\cH_x$.

The locality results we prove are expressed in terms of the support of local observables. 
Here, the support of an observable is understood as follows. If $X \subset \Gamma$, 
we identify $A \in \A_{X}$ with $A \otimes \idty_{\Gamma \setminus X} \in \A_{\Gamma}$. 
In a similar manner, we have that for each $X \subset Y \subset \Gamma$, $\A_{X} 
\subset \A_{Y}$. The support of an observable 
$A \in \A_{\Gamma}$ is then the minimal set $X \subset \Gamma$ for which $A =
A' \otimes \idty_{\Gamma \setminus X}$ with $A' \in \A_{X}$. 

The models we consider here correspond to bounded perturbations of
local self-adjoint Hamiltonians. Specifically, we will fix a collection of on-site local operators
$\{ H_x \}_{x \in \Gamma}$ where each $H_x$ is a self-adjoint
operator with domain $D(H_x) \subset \mathcal{H}_x$. In addition, we will consider a 
general class of bounded perturbations.
These are defined in terms of an interaction $\Phi$, which is a map from the
set of subsets of $\Gamma$ to $\mathcal{A}_{\Gamma}$ with the property that
for each set $X \subset \Gamma$, $\Phi(X) \in \mathcal{A}_X$ and
$\Phi(X) ^*= \Phi(X)$. 

Lieb-Robinson bounds are essentially an upper bound for the velocity with which
perturbations can propagate through the system. The metric $d$ measure distances
in the underlying space and it turns out that some regularity of $\Gamma$ has to be assumed,
which can be interpreted as a condition which guarantees that $\Gamma$ can be
`nicely' embedded in a finite-dimensional space. This property is expressed in terms of
a non-increasing, real-valued function $F: [0, \infty) \to (0, \infty)$, which will enter
in our estimate of the Lieb-Robinson velocity. $F$ will also be used to impose a
decay condition on the interactions in the system. The existence of a function $F$ with
the required properties is non-trivial only when the cardinality of $\Gamma$ is infinite, but
all the relevant quantities can be defined for a finite system. To each pair $x,y \in \Gamma$, 
there is a number $ \tilde{C}_{x,y}$ such that
\begin{equation} \label{eq:convf}
\sum_{z \in \Gamma} F \left( d(x,z) \right) F \left( d(z,y) \right) \leq \tilde{C}_{x,y} 
F \left( d(x,y) \right) \, .
\end{equation}
Take $C_{x,y}$ to be the infimum over all such $\tilde{C}_{x,y}$, and denote by 
$C = \max_{x,y \in \Gamma} C_{x,y}$.
Explicitly, for models with $\Gamma \subset \mathbb{Z}^d$, one choice of $F$ is given by 
$F(r) = (1+r)^{d + 1}$. In this case, the convolution constant 
may be taken as $C = 2^{d+1} \sum_{x \in \Gamma}F(|x|)$.
In general, the quantity 
\be
\| \Phi \| = \max_{x,y \in \Gamma} \sum_{\stackrel{X \subset \Gamma:}{x,y \in X}} 
\frac{\| \Phi(X) \|}{F \left( d(x,y) \right)} \, ,
\ee
which is finite for any interaction $\Phi$ over $\Gamma$, will also play a role in our analysis.
 
Now, for a fixed sequence of local Hamiltonians $\{H_x \}_{x\in\Gamma}$, as
described above, an interaction $\Phi$, and any subset $\Lambda \subset \Gamma$, 
we will consider self-adjoint Hamiltonians of the form
\begin{equation} \label{eq:localham}
H_{\Lambda} \, = \, H^{\rm loc}_{\Lambda} \, + \, H^{\Phi}_{\Lambda} \, = \, 
\sum_{x \in \Lambda} H_x \, + \, \sum_{X \subset \Lambda} \Phi(X),
\end{equation}
with domain $\bigotimes_{x \in \Lambda} D(H_x) \subset \mathcal{H}_{\Lambda}$. 
Since each operator $H_{\Lambda}$ is self-adjoint, it generates a 
Heisenberg dynamics, or time evolution, $\{ \tau_t^{\Lambda} \}$,
which is the one parameter group of automorphisms defined by
\begin{equation}
\tau_t^{\Lambda}(A) \, = \, e^{it H_{\Lambda}} \, A \, e^{-itH_{\Lambda}} 
\quad \mbox{for any} \quad A \in \mathcal{A}_{\Lambda}.
\end{equation}

Let us consider two concrete models of the type described above.

\begin{example} Take $\mathcal{H}_x = \C^2$ for each $x \in \Gamma$ and
consider a Heisenberg Hamiltonian of the form
\be
H_{\Gamma} = \sum_{x\in\Gamma} B S^3_x\ +\ 
\sum_{\stackrel{x,y \in \Gamma:}{d(x,y) \leq 1}} J_{xy}{\bf S}_x\cdot {\bf S}_y \, .
\ee
As we will see below, the velocity corresponding to such a model depends on the 
interaction strength
$J_{xy}$, and it is independent of the external magnetic field $B$.
\end{example}

\begin{example} \label{ex:osc}
Take $\mathcal{H}_x = L^2( \R, d q_x)$ for each $x \in \Gamma$ and
consider an oscillator Hamiltonian of the form
\be
H_{\Gamma} = \sum_{x\in\Gamma}  p_x^2 \ + \ V(q_x) +  \sum_{\stackrel{x,y \in \Gamma:}{d(x,y) 
\leq 1}} \phi(q_x-q_y) \, .
\ee
If $H_x = p_x^2 + V(q_x)$ is self-adjoint, then Theorem~\ref{thm:phi} below estimates the 
velocity of such a model for any real-valued $\phi \in L^{\infty}( \mathbb{R})$.
\end{example}

For a Hamiltonian that is a sum of local interaction terms, nearest neighbor for example,
it is reasonable to expect that the spread of the support of a time evolved observable
depends only on the surface area of the support of the observable being 
evolved; not the full volume of the support. This is important in some applications,
e.g., to prove the split property of gapped spin chains \cite{matsui:2010}.
To express the dependence on the surface area  we will use the following notation. 
Let $X \subset \Lambda \subset \Gamma$.
Denote the surface of $X$, regarded as a
subset of $\Lambda$, by
\begin{equation} \label{eq:defsurf}
S_{\Lambda}(X) \, = \, \left\{ Z \subset \Lambda \, : \, Z \cap X \neq \emptyset
\mbox{ and }  Z \cap X^c \neq \emptyset \right\} \, ,
\end{equation}
and we will write $S(X) =S_{\Gamma}(X)$. The $\Phi$-boundary of a set $X$, written 
$\partial_{\Phi} X$, 
is given by
\begin{equation}
\partial_{\Phi} X \, = \, \left\{ x \in X \, : \, \exists Z \in S(X)
  \mbox{ with } x \in Z \mbox{ and } \Phi(Z) \neq 0 \, \right\}.
\end{equation}
Denote the distance between two sets $X, Y \subset \Gamma$ by $d(X,Y) 
= \min_{x \in X, y \in Y}d(x,y)$.

The main result of this section is the following theorem, which was first proved in
\cite{nachtergaele:2009a}.

\begin{theorem}\label{thm:phi} Let $\Gamma$ be a finite set and fix a collection of local
 Hamiltonians $\{ H_x \}_{x \in \Gamma}$ and an interaction $\Phi$ over $\Gamma$.
 Let $X$ and $Y$ be subsets of $\Gamma$ with $d(X,Y) >0$ and take any set 
 $\Lambda \supset X \cup Y$.
 For any pair of local observables $A \in \mathcal{A}_X$ and $B \in \mathcal{A}_Y$, the estimate
\begin{equation} \label{eq:lrbd1}
\left\| [ \tau_t^{\Lambda}(A), B ] \right\| \, \leq \, \frac{2 \, \| A \|
\, \|B \|}{C} \, \left( e^{2 C  \| \Phi \|  |t|} -1 \right)   \, D(X,Y),
\end{equation}
holds for all $t \in \R$. Here
\begin{equation} \label{eq:defda}
D(X,Y) =
\min \left[ \sum_{x \in \partial_\Phi X} \sum_{y \in
  Y} \, F \left( d(x,y) \right), \sum_{x \in X} \sum_{ y \in
  \partial_\Phi Y} \, F \left( d(x,y) \right)\right].
\end{equation}
\end{theorem}

Before we prove Theorem~\ref{thm:phi}, we make a few comments
which may be useful in interpreting this result. 
First, we note that if $X$ and $Y$ have a non-empty
intersection, then the argument provided below produces an analogous bound with 
the factor $e^{2 \, \| \Phi \| \,C \, |t|} - 1 $ replaced by 
$ e^{2 \, \| \Phi \| \, C \, |t|}  $. In the case of empty intersection 
and for small values of $|t|$, (\ref{eq:lrbd1}) is a better and sometimes
more useful estimates than the obvious bound 
$\| [ \tau_t(A), B] \| \leq 2 \| A \| \| B \|$, valid for all 
$t \in \mathbb{R}$.

Next, (\ref{eq:lrbd1}) provides a fairly explicit locality estimate for the corresponding 
dynamics. We have thus far expressed this in terms of the function $F$ defined above. 
Note that if $\Gamma$ is equipped with such a function $F$, then for every 
$\mu >0$, $F_{\mu}(r) = e^{- \mu r} F(r)$ also satisfies (\ref{eq:convf}) above.  Setting then
\begin{equation}
\| \Phi \|_{\mu} = \max_{x,y \in \Gamma} \sum_{\stackrel{X \subset \Gamma:}{x,y \in X}} 
\frac{\| \Phi(X) \|}{F_{\mu} \left( d(x,y) \right)} \, ,
\end{equation}
 for all $\mu >0 $, it is easy to see that
\begin{equation} \label{eq:dbd}
D_\mu(X,Y) \, \leq \,  \, \| \Phi \|_\mu \, \min \left( \left| \partial_{\Phi}X \right|, \left| 
\partial_{\Phi}Y \right| \right) \, e^{-\mu \, d(X,Y)} \, \max_{y \in \Gamma} \sum_{x \in \Gamma} 
F(d(x,y)) \, .
\end{equation}
In this case, (\ref{eq:lrbd1}) implies
\begin{equation} \label{eq:vel}
\left\| [ \tau_t^{\Lambda}(A), B ] \right\| \, \leq \, \frac{2 \, \| A \|
\, \|B \|}{ C_\mu} \, \max_{y \in \Gamma} \sum_{x \in \Gamma} F(d(x,y)) \,  \min \left( \left|
    \partial_{\Phi}X \right|, \left| \partial_{\Phi}Y \right| \right) \, e^{- \mu \,\left[
 d(X,Y) - \frac{2 \| \Phi \|_\mu C_\mu}{\mu} |t| \right]},
\end{equation}
i.e. the locality bounds decay exponentially in space with arbitrary rate $\mu >0$.
For every $\mu$, the system's velocity of propagation, $v_{\Phi}$, satisfies the bound
\begin{equation}
v_{\Phi} \leq \frac{2 \| \Phi \|_\mu C_\mu}{\mu}.
\end{equation}

As a final comment, we observe that  for fixed local observables $A$ and $B$, the 
bounds above, (\ref{eq:lrbd1}) and (\ref{eq:vel}), are 
independent of the volume $\Lambda \subset \Gamma$; given that $\Lambda$
contains the supports of both $A$ and $B$. Furthermore, we note that  
these bounds place a constraint on only the minimum of the support of the
two observables. Thus the estimate is still independent of $\Lambda$ even if
the support of one of the observables depends on the volume $\Lambda$.

The proof of Theorem~\ref{thm:phi} uses a basic 
lemma about the growth of the solutions of first order, 
inhomogeneous differential equations. We state and prove it
before the proof of the theorem.

Let $\mathcal{B}$ be a Banach space. For each $t \in \R$, 
let $A( t) : \mathcal{B} \to \mathcal{B}$ be a linear
operator, and denote by $X( t)$ the solution of the 
differential equation 
\begin{equation} \label{eq:fode}
\partial_{t} X( t) \, = \, A( t) \, X( t)
\end{equation}
with boundary condition $X(0) = x_0 \in \mathcal{B}$. We say that the
family of operators $A(t)$ is {\em norm-preserving} if for 
every $x_0 \in \mathcal{B}$, the mapping $\gamma_{t} :
\mathcal{B} \to \mathcal{B}$ which associates $x_0 \mapsto X( t)$,
i.e., $\gamma_{t}(x_0) = X( t)$, satisfies
\begin{equation} \label{eq:normp}
\| \, \gamma_{t}(x_0) \, \| \, = \, \| \, x_0 \, \| \quad \mbox{for all } t \in
\R.
\end{equation}

Some obvious examples are the case where $\mathcal{B}$ is a Hilbert space 
and $A(t)$ is anti-hermitian for each $t$, or when $\mathcal{B}$ 
is a $*$-algebra of operators on a Hilbert space with a spectral norm and,
for each $t$, $A(t)$ is a derivation commuting with the $*$-operation.

\begin{lemma} \label{lem:normp} Let $A( t)$, for $t \in \R$, be a family of 
norm preserving operators in some Banach space $\mathcal{B}$. For any
function $B : \R \to \mathcal{B}$, the solution of 
\begin{equation} \label{eq:inhom}
\partial_{t} Y( t) \, = \, A( t) Y( t) \, + \, B( t),
\end{equation}
with boundary condition $Y(0) = y_0$, satisfies the bound
\begin{equation} \label{eq:yest}
\| \, Y( t) \, - \, \gamma_{t}(y_0) \, \| \, \leq \, \int_0^{ t}  \| \, B( t')
\, \| \, d t' .
\end{equation}
\end{lemma}

\begin{proof}
For any $t \in \R$, let $X( t)$ be the solution of 
\begin{equation} \label{eq:fode1}
\partial_{t} X( t) \, = \, A( t) \, X( t)
\end{equation}
with boundary condition $X(0) = x_0$, and let $\gamma_{t}$ be the
linear mapping which takes $x_0$ to $X( t)$. By variation of constants,
the solution of the inhomogeneous equation (\ref{eq:inhom}) may be
expressed as
\begin{equation} \label{eq:ysol}
Y( t) \, = \, \gamma_{t} \left( \, y_0 \, + \, \int_0^{
      t} ( \gamma_s)^{-1} \left( B(s) \right) ds \, \right).
\end{equation}
The estimate (\ref{eq:yest}) follows from (\ref{eq:ysol}) as $A( t)$ is
norm preserving.
\end{proof}

\begin{proof}[Proof of Theorem \ref{thm:phi}]
Fix $\Lambda \subset \Gamma$ as in the statement of the theorem. 
As this set will remain fixed throughout the argument, we will suppress it in our notation.
In particular, we will denote $\tau_t^{\Lambda}$ merely by $\tau_t$. 

Without loss of generality, we will assume that 
\begin{equation}
D(X,Y) = \sum_{x \in \partial_{\Phi}X} \sum_{y \in Y} F \left( d(x,y) \right) \, .
\end{equation}
Otherwise, we apply the argument below to $\| [ \tau_{-t}(B), A ] \|$.

For each $Z \subset \Gamma$, we introduce the quantity
\begin{equation} \label{def:cbxt}
C_B(Z; t) \, := \, \sup_{A \in \mathcal{A}_Z, A\neq 0} \frac{ \| [ \tau_t(A), B
  ] \| }{ \| A \|},
\end{equation}
and note that $C_B(Z; 0) \leq 2 \|B \| \delta_Y(Z)$, where we defined $\delta_Y (Z) = 1$ if 
$Y\cap Z \neq \emptyset$ and $\delta_Y (Z) = 0$ if $ Y\cap Z = \emptyset$.
Note that the dynamics generated by
\[ H_{\Lambda}^{\text{loc}} \, + \, H_X^{\Phi} \, = \, \sum_{x \in \Lambda} H_x \, + \, 
\sum_{Z \subset X} \Phi(Z) \] 
leaves $\mathcal{A}_X$ invariant. More precisely, if we define $\tau^{\rm loc}_t$ by
\begin{equation}\label{eq:tauloc}
\tau_t^{\rm{loc}}( A) \, = \, e^{it \left(H^{\rm loc}_{\Lambda} + H_X^{\Phi} \right)} \, A \, 
e^{-it \left(H^{\rm loc}_{\Lambda} + H_X^{\Phi} \right)} \quad \mbox{for all} \quad A \in 
\mathcal{A}_{\Lambda},
\end{equation}
then we have that for every $A \in \cA_X$, $\tau_t^{\rm{loc}} (A) \in \cA_X$ for all $t\in \bR$. 
This implies, recalling the definition (\ref{def:cbxt}), that
\begin{equation}
C_B(X; t) \, = \, \sup_{A \in \mathcal{A}_X, A\neq 0} \frac{ \| [ \tau_t^{\Lambda}
(\tau_{-t}^{\text{loc}} (A)), B ] \| }{ \| A \|}\,.
\label{eq:cbx}\end{equation}

Consider the function 
\begin{equation}
f(t) \, := \, \left[ \tau_t \left( \tau_{-t}^{\rm{loc}}(A) \right), B \right],
\end{equation}
for $A \in \mathcal{A}_X$, $B \in \mathcal{A}_Y$, and $t \in
\mathbb{R}$. It is straightforward to verify that
\begin{equation} \label{eq:derf}
f'(t) = i \sum_{Z \in S_{\Lambda}(X)} \left[ \tau_t \left( \Phi(Z)
    \right), f(t) \right] - i  \sum_{Z \in \, S_{\Lambda}(X)} \left[
    \tau_t( \tau_{-t}^{{\rm loc}}(A)), \left[ \tau_t \left( \Phi(Z) \right), B \right] \right] .
\end{equation}
The first term in the above differential
equation is norm preserving, see Lemma~\ref{lem:normp}, and therefore we have the bound
\begin{equation} \label{eq:normpresbd}
\| f(t) \| \, \leq \, \| f(0) \| \, + \, 2 \| A \| \sum_{Z \in S(X)} \int_0^{|t|} \| [ \tau_s(\Phi(Z)), B ] \| ds.
\end{equation}
Recalling definition (\ref{def:cbxt}), the above inequality readily
implies that
\begin{equation} \label{eq:recurbd}
C_B(X,t) \leq C_B(X,0) + 2 \sum_{ Z \in \, S(X)} \| \Phi(Z) \| \int_0^{|t|} C_B(Z, s) ds,
\end{equation}
where we have used (\ref{eq:cbx}).
Iteration of (\ref{eq:recurbd}) yields that 
\begin{equation}  \label{eq:seriesbd}
C_B(X,t) \, \leq \, 2 \| B \| \, \sum_{n=1}^{ \infty}
\frac{(2|t|)^n}{n!} a_n,
\end{equation}
where for $n \geq 1$,
\begin{equation}
a_n \, = \, \sum_{Z_1 \in S(X)} \sum_{Z_2 \in S(Z_1)} \cdots \sum_{Z_n
  \in S(Z_{n-1})} \delta_Y(Z_n) \, \prod_{i=1}^n \| \Phi(Z_i) \| \, .
\end{equation}

For any interaction $\Phi$, one may estimate that
\begin{equation} 
a_1 \, \leq \, \sum_{y \in Y} \sum_{\stackrel{Z \in
    S(X):}{y \in Z}} \| \Phi(Z) \| \, \leq \, \| \Phi \| \sum_{y \in Y}
\sum_{x \in \partial_{\Phi}X} F \left( d(x, y) \right).
\end{equation}
In addition,
\begin{eqnarray} 
a_2 & \leq & \sum_{y \in Y} \sum_{Z_1 \in S(X)} 
\| \Phi(Z_1) \| \sum_{z_1 \in \partial_{\Phi}Z_1} 
\sum_{ \stackrel{Z_2 \subset \Gamma:}{z_1, y \in Z_2}}  \| \Phi(Z_2) \|  \nonumber \\ 
& \leq &   \| \Phi \| \, \sum_{y \in Y} \sum_{z_1 \in
  \Gamma} F \left( d(z_1, y) \right)  \, \sum_{ \stackrel{Z_1 \in S(X):}{z_1 \in Z_1}} 
\| \Phi(Z_1) \|  \nonumber \\ & \leq &   \| \Phi \|^2 \, \sum_{x \in \partial_{\Phi}X} \sum_{y \in Y} 
\sum_{z_1 \in
  \Gamma} F \left( d(x,z_1) \right) \, F \left( d(z_1, y) \right) \nonumber \\
 & \leq & \| \Phi \|^2 \, C \, \sum_{x \in \partial_{\Phi} X} \sum_{y \in
  Y} F \left( d(x,y) \right),
\end{eqnarray}
where we have used $C$ from (\ref{eq:convf}) for the final inequality. 
With analogous arguments, one finds that for all $n \geq 1$,
\begin{equation} \label{eq:aneq}
a_n \, \leq \,  \| \Phi \|^n \, C^{n-1} \, \sum_{x \in \partial_{\Phi}X} \sum_{y \in
  Y} F \left( d(x,y) \right).
\end{equation}
Inserting (\ref{eq:aneq}) into (\ref{eq:seriesbd}) we see that
\begin{equation} \label{eq:lrbdd}
C_B(X,t) \, \leq \,  \frac{2 \, \| B \| }{C} \, \left( e^{2 C \| \Phi \|  |t|} - 1
\right) \, \sum_{x \in \partial_{\Phi}X} \sum_{y \in Y} \, F \left( d(x,y) \right),
\end{equation}
from which (\ref{eq:lrbd1}) immediately follows. 
\end{proof}

%
%
%
%
%
%

\subsection{Unbounded Interactions}\label{sec:unbounded}

In this section we will consider locality estimates for systems with 
unbounded interaction terms. There are very few results in this
context. Results bounding the speed of propagation of perturbations
in classical anharmonic lattice systems have been obtained \cite{marchioro:1978,
marchioro:1981,butta:2007}, but these works do not
provide explicit estimates for the Lieb-Robinson velocity.
For a class of classical models similar to the
quantum models we will discuss here, bounds for the Lieb-Robinson velocity have
been proved recently in \cite{raz:2009}.  The only known results for quantum systems
in this context apply to harmonic systems and a class of bounded perturbations and were
obtained in \cite{nachtergaele:2009a} and \cite{nachtergaele:2010}. These are the
results we will review here.

We begin in Section~\ref{sec:harmham} by introducing a well-known family of
harmonic oscillator models defined on finite subsets of $\mathbb{Z}^d$; compare with
Example~\ref{ex:osc}. Then, in Section~\ref{sec:weylop}, we introduce a
convenient class of observables, the Weyl operators, which the harmonic dynamics 
leaves invariant. In Section~\ref{sec:harmlrbfv}, we demonstrate a Lieb-Robinson bound for
these harmonic models. Finally, in Section~\ref{subsec:anharmlrbs}, we show that a
similar Lieb-Robinson bound holds for a large class of bounded perturbations.

\subsubsection{Harmonic Oscillators} \label{sec:harmham}

We first consider a system of coupled harmonic oscillators restricted to a finite volume. 
Specifically on cubic subsets $\Lambda_L \, = \, \left( -L, L \right]^d  \subset \mathbb{Z}^d$,
we analyze Hamiltonians of the form
\begin{equation} \label{eq:harham}
H_L^{h} \, = \,  \sum_{ x \in \Lambda_L} p_{ x }^2 \, +\, \omega^2 \, q_{ x}^2 \, + \,
\sum_{j = 1}^{d}  \lambda_j \, (q_{ x } - q_{ x + e_j})^2
\end{equation}
acting in the Hilbert space
\begin{equation} \label{eq:hspace}
\mathcal{H}_{\Lambda_L} = \bigotimes_{x \in \Lambda_L} L^2(\mathbb{R}, dq_x).
\end{equation}
Here the quantities $p_x$ and $q_x$, which appear in (\ref{eq:harham}) above, are the 
single site momentum and position operators regarded as operators on the full Hilbert space 
$\mathcal{H}_{\Lambda_L}$
by setting 
\begin{equation} \label{eq:pandq}
p_x = \idty \otimes \cdots \otimes \idty
\otimes -i \frac{d}{dq} \otimes \idty \cdots \otimes \idty \quad
\mbox{ and } \quad q_x = \idty \otimes \cdots \otimes \idty \otimes q \otimes \idty
\cdots \otimes \idty,
\end{equation}
i.e., these operators act non-trivially only in the $x$-th factor of $\mathcal{H}_{\Lambda_L}$. 
These operators satisfy the canonical commutation relations
\begin{equation} \label{eq:comm}
[p_x, p_y] \, = \, [q_x, q_y] \, = \, 0 \quad \mbox{ and } \quad
[q_x, p_y] \, = \, i \delta_{x,y},
\end{equation}
valid for all $x, y \in \Lambda_L$.  In addition,  $\{ e_j \}_{j=1}^{d}$ are the canonical basis
vectors in $\mathbb{Z}^{d}$, the numbers
$\lambda_j \geq 0$ and $\omega > 0$ are the parameters of the system, and the Hamiltonian
is assumed to have periodic boundary conditions, in the sense that $q_{x+e_j} = q_{x-(2L-1)e_j}$ 
if $x \in \Lambda_L$ but $x+ e_j \not\in \Lambda_L$. It is well-known that these Hamiltonians
are essentially self-adjoint on $C_0^{\infty}$, see e.g \cite{reed:1975}. Moreover, these operators
have a diagonal representation in Fourier space. We review this quickly to establish some
notation and refer the interested reader to \cite{nachtergaele:2009a} for more details.
 
Introduce the operators
\begin{equation} \label{eq:Q+Pk}
Q_k \, = \, \frac{1}{ \sqrt{ | \Lambda_L |}} \sum_{x \in \Lambda_L} e^{- i k \cdot x} q_x \quad 
\mbox{and} \quad
P_k \, = \, \frac{1}{ \sqrt{ | \Lambda_L |}} \sum_{x \in \Lambda_L} e^{- i k \cdot x} p_x \, ,
\end{equation}
defined for each $k \in \Lambda_L^* \, = \, \left\{ \, \frac{x \pi}{L} \, : \, x \in \Lambda_L \, \right\} $,
and set 
\begin{equation} \label{eq:defgamma}
\gamma(k) \, = \,  \sqrt{ \omega^2 \, + \, 4 \sum_{j=1}^{d} \lambda_j \, \sin^2(k_j/2) }.
\end{equation}
A calculation shows that 
\begin{equation} \label{eq:diagham}
H_L^h \, = \, \sum_{k \in \Lambda_L^*} \, \gamma(k) \, \left( \, 2 \, b_k^*\,  b_k \, + \, 1 \, \right) \, 
\end{equation}
where the operators $b_k$ and $b_k^*$ satisfy
\begin{equation} \label{eq:beqns}
b_k \, = \, \frac{1}{ \sqrt{2 \gamma(k)}} \, P_k - i \sqrt{ \frac{\gamma(k)}{2}} \, Q_k 
\quad {\rm and} \quad
b_k^* \, = \, \frac{1}{ \sqrt{2 \gamma(k)}} \, P_{-k} + i \sqrt{ \frac{\gamma(k)}{2}} \, Q_{-k} \, .
\end{equation}
In this sense, we regard the Hamiltonian $H_L^h$ as diagonalizable. The special case of
$\omega =0$ is discussed in \cite{nachtergaele:2009a}.

\subsubsection{Weyl Operators} \label{sec:weylop}

For these harmonic Hamiltonians, a specific class of observables, namely the Weyl operators, 
is particularly convenient.
Given any function $f: \Lambda_L \to \mathbb{C}$, the corresponding Weyl operator $W(f)$ 
is defined by setting
\begin{equation}
W(f) = \mbox{exp} \left[ i \sum_{x \in \Lambda_L} \mbox{Re}[f(x)] q_x + \mbox{Im}[f(x)] p_x \right] \, .
\end{equation}
It is easy to see that each $W(f)$ is a unitary operator with
\begin{equation}
W(f)^{-1} = W(-f) = W(f)^* \, .
\end{equation}
Moreover, using the well-known Baker-Campbell-Hausdorff formula
\begin{equation}
e^{A+B} = e^Ae^Be^{-[A,B]/2} \quad \mbox{if } [A,[A,B]] =[B, [A,B]] = 0 \, ,
\end{equation}
and the commutation relations (\ref{eq:comm}), it follows that Weyl operators satisfy
the Weyl relations
\begin{equation} \label{eq:weylrel}
W(f) W(g) = W(g) W(f) e^{-i \mbox{Im}[ \langle f, g \rangle ]} = W(f+g) e^{-i 
\mbox{Im}[ \langle f, g \rangle ] /2} \, ,
\end{equation}
for any $f,g : \Lambda_L \to \mathbb{C}$. These operators also generate shifts of the position 
and momentum operators in the sense that
\begin{equation}
W(f)^* q_x W(f) = q_x - \mbox{Im}[ f(x)] \quad \mbox{and} \quad W(f)^* p_x W(f) = p_x + 
\mbox{Re}[f(x)] \, .
\end{equation}
The algebra generated by all such Weyl operators is called the Weyl algebra.

A key observation which we will exploit in our locality estimates is the fact that the harmonic 
dynamics leaves the Weyl algebra invariant. We state this as a lemma.
Fix $L \geq 1$ and $t \in \mathbb{R}$. Denote by $\tau_t^{h, L}$ the harmonic dynamics 
generated by $H_L^h$, i.e. for any $A \in \mathcal{B}( \mathcal{H}_{\Lambda_L})$ set
\begin{equation}
\tau_t^{h,L}(A) = e^{itH_L^h} A e^{-itH_L^h} \, .
\end{equation}

\begin{lemma} \label{lem:weylinv} Fix $L \geq 1$ and $t \in \mathbb{R}$. There exists a mapping 
$T^{h,L}_t : \ell^2( \Lambda_L) \to \ell^2( \Lambda_L)$ for which
\begin{equation}
\tau_t^{h,L}( W(f)) = W( T_t^{h,L} f ) \, 
\end{equation}
for any $f \in \ell^2( \Lambda_L)$.
\end{lemma}

\begin{proof}
Since we will fix $L \geq 1$ throughout this argument and only consider harmonic Hamiltonians, 
we will denote by $\tau_t = \tau_t^{h,L}$ and $T_t = T_t^{h,L}$ to ease notation.
To prove this lemma, it is also convenient to express a given Weyl operator in terms of 
annihilation and creation operators, i.e.,
\begin{equation} \label{eq:anncre}
a_x \, = \, \frac{1}{\sqrt{2}} \left( q_x \, + \, i p_x \right) \quad \mbox{and} \quad a^*_x \, = \, 
\frac{1}{\sqrt{2}} \left( q_x \, - \, i p_x \right),
\end{equation}
which satisfy 
\begin{equation} \label{eq:comrela}
[a_x, a_y] = [a_x^*, a_y^*] = 0 \quad \mbox{and} \quad [a_x, a_y^*] = \delta_{x,y} \quad 
\mbox{for all } x,y \in \Lambda_L \, .
\end{equation}
One finds that
\begin{equation} \label{eq:weyla}
W(f)  =  \mbox{exp} \left[ \frac{i}{\sqrt{2}} \left( a(f) \, + \, a^*(f) \right) \right] \, 
\end{equation}
where, for each $f \in \ell^2(\Lambda_L)$, we have set
\begin{equation} \label{eq:defafa*f}
a(f) \, = \, \sum_{x \in \Lambda_L} \overline{f(x)} \, a_x,  \quad a^*(f) \, = \, 
\sum_{x \in \Lambda_L} f(x) \, a_x^*\, .
\end{equation}

It is easy to see that the harmonic dynamics acts trivially on the diagonalizing operators $b$, i.e.,
\begin{equation}
\tau_t(b_k) = e^{-2i \gamma(k) t} b_k \quad \mbox{and} \quad \tau_t(b_k^*) = e^{2i 
\gamma(k) t} b_k^* \, , 
\end{equation}
where $b_k$ and $b_k^*$ are as defined in (\ref{eq:beqns}). 
Hence, if we further introduce
\begin{equation} \label{eq:defb}
b_x = \frac{1}{ \sqrt{| \Lambda_L|}} \sum_{k \in \Lambda_L^*} e^{ikx} b_k \quad \mbox{and} \quad 
b_x^* = \frac{1}{ \sqrt{| \Lambda_L|}} \sum_{k \in \Lambda_L^*} e^{ikx} b_k^*, 
\end{equation}
for each $x \in \Lambda_L$ and, analogously to (\ref{eq:defafa*f}), define
\begin{equation} \label{eq:defbfb*f}
b(f) \, = \, \sum_{x \in \Lambda_L} \overline{f(x)} \, b_x,  \quad b^*(f) \, = \, 
\sum_{x \in \Lambda_L} f(x) \, b_x^*,
\end{equation}
for each $f \in \ell^2( \Lambda_L)$, then one has that
\begin{equation} \label{eq:taub}
\tau_t \left( b(f) \right) = b \left( [\mathcal{F}^{-1} M_t \mathcal{F}] f \right) \, , 
\end{equation}
where $\mathcal{F}$ is the unitary Fourier transform  from $\ell^2(\Lambda_L)$ to 
$\ell^2(\Lambda_L^*)$and $M_t$ is the 
operator of multiplication by $e^{2i \gamma(k) t}$ in Fourier space with $\gamma(k)$ as in 
(\ref{eq:defgamma}). 
The proof is completed by demonstrating a change of variables relation between the $a$'s 
and the $b$'s.

A short calculation shows that there exists a linear mapping $U: \ell^2( \Lambda_L) 
\to \ell^2( \Lambda_L)$
and an anti-linear mapping $V: \ell^2( \Lambda_L) \to \ell^2( \Lambda_L)$ for which
\begin{equation} \label{eq:b=a}
b(f) = a(Uf) + a^*(Vf) \, ,
\end{equation}
a relation know in the literature as a Bogoliubov transformation \cite{manuceau:1968}.
In fact, one has that
\begin{equation} \label{eq:defU+V}
U = \frac{i}{2} \mathcal{F}^{-1} M_{\Gamma_+} \mathcal{F} \quad \mbox{ and } \quad V = 
\frac{i}{2} \mathcal{F}^{-1} M_{\Gamma_-} \mathcal{F} J
\end{equation}
where $J$ is complex conjugation and $M_{\Gamma_{\pm}}$ is the operator of multiplication
by
\begin{equation} \label{eq:multg}
\Gamma_{\pm}(k) = \frac{1}{\sqrt{ \gamma(k)}} \pm \sqrt{ \gamma(k)} \, ,
\end{equation}
again, with $\gamma(k)$ as in (\ref{eq:defgamma}).
Using the fact that $\Gamma_{\pm}$ is real valued and even, it is easy to check that
\begin{equation} \label{eq:bog1}
U^* U - V^* V = \idty = U U^* - V V^*
\end{equation}
and
\begin{equation} \label{eq:bog2}
V^* U - U^* V = 0 =  V U^* - UV^* \,
\end{equation}
where we stress that $V^*$ is the adjoint of the {\it anti-linear} mapping $V$. The relation 
(\ref{eq:b=a}) is invertible, in fact,
\begin{equation}
a(f) = b(U^*f) - b^*(V^*f) \, ,
\end{equation}
and therefore
\begin{equation}
W(f) = \mbox{exp} \left[  \frac{i}{ \sqrt{2}} \left( b((U^*-V^*)f) + b^*((U^*-V^*)f) \right) \right] \, .
\end{equation}
Clearly then,
\begin{equation}
\tau_t(W(f)) = W( T_tf) \, ,
\end{equation}
where the mapping $T_t$ is given by
\begin{equation}
T_t = (U+V) \mathcal{F}^{-1} M_t \mathcal{F} (U^*-V^*) \, ,
\end{equation}
and we have used (\ref{eq:taub}).
\end{proof}

\subsubsection{Lieb-Robinson bounds for harmonic Hamiltonians}
\label{sec:harmlrbfv}

In this section, we demonstrate a Lieb-Robinson type bound for these
harmonic lattice systems. We state our estimate, first proved in
\cite{nachtergaele:2009a}, as follows.

\begin{theorem} \label{thm:harmlrb} Fix $L \geq 1$. For any $\mu >0$, the estimate
\begin{equation}\label{eq:lrbharm}
\left\| \left[ \tau_t^{h,L} \left( W(f) \right), W(g) \right] \right\| \, \leq \,
C  \, \sum_{x, y \in \Lambda_L} \, |f(x)| \, |g(y)| \,
e^{-\mu \left( d(x,y) - c_{\omega,\lambda} \max \left( \frac{2}{\mu} \, , \, e^{(\mu/2)+1}\right) |t| \right)}
\end{equation}
holds for all functions $f,g \in \ell^2( \Lambda_L)$ and any $t \in \mathbb{R}$ . Here
\begin{equation}
\label{eq:dXY} d(x,y) = \sum_{j=1}^{d} \min_{\eta_j \in \, \bZ} |x_j-y_j + 2L \eta_j| \,.
\end{equation}
is the distance on the torus. Moreover
\begin{equation} \label{eq:defk}
C = \left(1+ c_{\omega,\lambda} e^{\mu/2} + c^{-1}_{\omega,\lambda} \right)
\end{equation} with $c_{\omega,\lambda} = (\omega^2 + 4 \sum_{j=1}^{\nu} \lambda_j)^{1/2}$.
\end{theorem}

Before we prove this result, we make a few comments. First, we denote by
\begin{equation}
v_h( \mu) = c_{\omega,\lambda} \max \left( \frac{2}{\mu} \, , \, e^{(\mu/2)+1}\right)
\end{equation}
our estimate on the harmonic Lieb-Robinson velocity corresponding to decay rate 
$\mu >0$. Optimizing over
$\mu >0$ produces a rate $1/2 < \mu_0 < 1$ for which $v_h( \mu_0) \leq 4 c_{\omega, \lambda}$.
Next, in \cite{nachtergaele:2009a}, see also \cite{nachtergaele:2010}, it is shown that the
mapping $T_t^{h,L}$
appearing in Lemma~\ref{lem:weylinv} can be expressed as a convolution. In fact, dropping 
the dependency on $h$ and $L$, one has that
\begin{equation} \label{eq:defft}
T_tf = f * \left(h_t^{(0)} - \frac{i}{2}(h_t^{(-1)} + h_t^{(1)}) \right) 
+ \overline{f}*\left( \frac{i}{2}(h_t^{(1)} - h_t^{(-1)}) \right).
\end{equation}
where
\begin{equation}\label{eq:h}
\begin{split}
h^{(-1)}_t(x) &= {\rm Im} \left[ \frac{1}{| \Lambda_L|} \sum_{k \in \Lambda_L^*} 
\frac{1}{\gamma(k)} \, e^{ik \cdot x - 2i \gamma(k) t} \, \right] \, ,
\\
h^{(0)}_t(x) &= {\rm Re} \left[  \frac{1}{| \Lambda_L|} \sum_{k \in \Lambda_L^*} 
e^{ik \cdot x - 2i \gamma(k) t} \,\right] \, ,
\\
h^{(1)}_t(x) &=  {\rm Im} \left[ \frac{1}{| \Lambda_L|} \sum_{k \in \Lambda_L^*} \gamma(k) \, 
e^{ik \cdot x - 2i \gamma(k) t} \, \right] \, .
\end{split}
\end{equation}
Next, by direct calculation, the following is proven in \cite{nachtergaele:2009a}.
\begin{lemma}\label{lem:htx}
Consider the functions defined in (\ref{eq:h}). For $\omega\geq 0, \lambda_1,
\ldots,\lambda_d\geq 0$, but such that  
$c_{\omega,\lambda} = (\omega^2 + 4 \sum_{j=1}^d \lambda_j )^{1/2} >0$, 
and any $\mu >0$, the bounds
\begin{equation}
\begin{split}
\left| h_t^{(0)}(x) \right| &\leq  e^{-\mu \left( |x| - c_{\omega,\lambda} \max \left( \frac{2}{\mu} \, , \, 
e^{(\mu/2)+1}\right) |t| \right)}
\\
\left| h_t^{(-1)}(x) \right| &\le  c^{-1}_{\omega,\lambda}e^{-\mu \left( |x| - c_{\omega,\lambda} 
\max \left( \frac{2}{\mu} \, , \, e^{(\mu/2)+1}\right) |t| \right)}
\\
\left| h_t^{(1)}(x) \right| &\le c_{\omega,\lambda}e^{\mu/2}e^{-\mu \left( |x| - c_{\omega,\lambda} 
\max \left( \frac{2}{\mu} \, , \, e^{(\mu/2)+1}\right) |t| \right)}
\end{split}
\end{equation}
hold for all $t \in \mathbb{R}$ and $x \in \Lambda_L$. Here  $|x| = \sum_{j=1}^{d} |x_i|$.
\end{lemma}

\begin{proof}[Proof of Theorem~\ref{thm:harmlrb}]
With $L \geq 1$ fixed, we again drop the
dependence of the dynamics on $h$ and $L$. Observe that
\begin{eqnarray}
\left[ \tau_t(W(f)), W(g) \right]  & = & \left\{ W( T_t f) - W(g) W( T_tf) W(-g) \right\} W(g)  \nonumber \\
& = & \left\{ 1 - e^{i \mbox{Im}[ \langle T_tf, g \rangle ]} \right\} W( T_tf) W(g) \, ,
\end{eqnarray}
where we have used Lemma~\ref{lem:weylinv} and the Weyl relations (\ref{eq:weylrel}).  
Since Weyl operators are unitary, the estimate
\begin{equation}
\left\|  \left[ \tau_t(W(f)), W(g) \right] \right\| \leq \left| \mbox{Im} \left[ \langle T_t f, g 
\rangle \right] \right|
\end{equation}
readily follows. The bound
\begin{equation}
\left| \mbox{Im} \left[ \langle T_t f, g \rangle \right] \right| \leq \sum_{y \in \Lambda_L} | 
T_tf(y) | \, |g(y) |
\end{equation} 
is obvious and (\ref{eq:lrbharm}) is now a consequence of (\ref{eq:defft}) and Lemma~\ref{lem:htx}. 
\end{proof}

In analogy with Section~\ref{subsec:bdint}, these Lieb-Robinson bounds can be
expressed in terms of a family of non-increasing, real-valued functions $F_{\mu} : [0, \infty) 
\to (0, \infty)$,
parametrized by $\mu>0$, given by 
\begin{equation}
F_{\mu}(r) = \frac{e^{- \mu r}}{(1+r)^{d+1}} \, .
\end{equation}
With $d( \cdot, \cdot)$ the metric on the torus, see (\ref{eq:dXY1}), it is easy to see that
\begin{equation} \label{eq:fconv}
\sum_{z \in \Lambda_L} F_{\mu}(d(x,z)) F_{\mu}(d(z,y)) \leq C_d \, F_{\mu}(d(x,y)) \, 
\end{equation}
with
\begin{equation}
C_d = 2^{d+1} \sum_{z \in \Lambda_L} \frac{1}{(1+|z|)^{d+1}} \, .
\end{equation}
The following corollary of Theorem~\ref{thm:harmlrb} is immediate.
\begin{corollary} \label{cor:harmlrbf}
Fix  $L \geq 1$. For any $\mu >0$ and $\epsilon >0$,
the estimate
\begin{equation}
\left\| \left[ \tau_t^{h,L} \left( W(f) \right), W(g) \right] \right\| \, \leq \,
C( \epsilon, \mu) e^{(\mu + \epsilon)v_h( \mu + \epsilon) | t|}  \, \sum_{x, y \in \Lambda_L} \, 
|f(x)| \, |g(y)| \, F_{\mu}(d(x,y))
\end{equation}
holds for all functions $f,g \in \ell^2( \Lambda_L)$ and any $t \in \mathbb{R}$. Here
\begin{equation}
\label{eq:dXY1} d(x,y) = \sum_{j=1}^{d} \min_{\eta_j \in \, \bZ} |x_j-y_j + 2L \eta_j| \,.
\end{equation}
is the distance on the torus,
\begin{equation}
C(\epsilon, \mu) = \left(1+ c_{\omega,\lambda} e^{(\mu+ \epsilon)/2} + 
c^{-1}_{\omega,\lambda} \right) \sup_{s\geq0} e^{- \epsilon s} ( 1 + s )^{d+1} \, ,
\end{equation} 
$c_{\omega,\lambda} = (\omega^2 + 4 \sum_{j=1}^{\nu} \lambda_j)^{1/2}$, and $v_h( \mu) = 
c_{\omega, \lambda} \max( \frac{2}{ \mu}, e^{\mu/2 +1})$
is the harmonic velocity corresponding to the decay rate $\mu$.
\end{corollary}

\subsubsection{Lieb-Robinson bounds for anharmonic systems} \label{subsec:anharmlrbs}

In this section we will consider perturbations of the harmonic Hamiltonians defined
in Section~\ref{sec:harmham}. The results we prove here originally appeared in 
\cite{nachtergaele:2009a} and \cite{nachtergaele:2010}.
The perturbations are defined as follows.
Fix $L \geq 1$. To each site $x \in \Lambda_L$, we will associate a 
finite measure $\mu_x$ on $\mathbb{C}$ and
an element $V_x \in \mathcal{B}( \mathcal{H}_{\Lambda_L})$ with the form
\begin{equation} \label{eq:defvx}
V_x = \int_{\mathbb{C}} W(z \delta_x) \mu_x(dz) \, .
\end{equation}
Here, for each $z \in \mathbb{C}$, $W(z \delta_x)$ is a Weyl operator as discussed in 
Section~\ref{sec:weylop}.
We require that each $\mu_x$ is even, i.e. invariant under $z \mapsto -z$, to ensure 
self-adjointness; namely $V_x^* = V_x$. The anharmonic Hamiltonians we consider
are given by
\begin{equation} \label{eq:defanharmham}
\begin{split}
H_L &= H_L^h + V \\
&= \sum_{x \in \Lambda_L} p_x^2 + \omega^2 q_x^2 + \sum_{j=1}^d 
\lambda_j (q_x -q_{x+e_j})^2 + \sum_{x \in \Lambda_L} V_x \, .
\end{split}
\end{equation}
We denote the dynamics generated by $H_L$ on $\mathcal{B}( \mathcal{H}_{ \Lambda_L})$ 
as $\tau_t^L$, that is
\begin{equation}
\tau_t^L(A) = e^{itH_L} A e^{-itH_L} \quad \mbox{for } A \in 
\mathcal{B}( \mathcal{H}_{ \Lambda_L}) \, .
\end{equation}

Before we present our Lieb-Robinson bounds, we discuss two examples.
\begin{example} Let $\mu_x$ be supported on $[- \pi, \pi)$ and absolutely 
continuous with respect to Lebesgue measure, i.e. 
$\mu_x(dz) = v_x(z) dz$. If $v_x$ is in $L^2([-\pi,\pi))$, then $V_x$ is proportional to an 
operator of multiplication by the 
inverse Fourier transform of $v_x$. Moreover, since the support of $\mu_x$ is real, $V_x$ 
corresponds to multiplication by a function depending only on $q_x$.
\end{example}
\begin{example}
Let $\mu_x$ have finite support, e.g., take $\mbox{supp}( \mu_x) = \{ z, -z \}$ for some number
$z = \alpha +  i \beta \in \mathbb{C}$. Then
\begin{equation}
V_x = W( z \delta_x) + W(- z \delta_x) = 2 \cos( \alpha q_x + \beta p_x) \, .
\end{equation}
\end{example}

We now state our first result.

\begin{theorem} \label{thm:ahlrb}
Fix $L \geq 1$ and take $V$, $H_L$, and $\tau_t^L$ as defined above. Suppose that
\begin{equation}
\kappa = \sup_{x \in \Lambda_L} \int_{\mathbb{C}} |z|^2 | \mu_x| (d z) < \infty \, .
\end{equation}
For every $\mu >0$ and $\epsilon >0$, there exist positive numbers $c$ and $v$ for which
the estimate
\begin{equation} \label{eq:anharmbd}
\left\| \left[ \tau_t^L \left( W(f) \right), W(g) \right] \right\| \leq c e^{ (v + c \kappa C_d ) |t|} 
\sum_{x, y \in \Lambda_L} |f(x)| \, |g(y)| F_{\mu} \left( d(x,y) \right)
\end{equation}
holds for all functions $f, g \in \ell^2( \Lambda_L)$ and any $t \in \mathbb{R}$. Here
\begin{equation}
c = C( \epsilon, \mu) \quad \mbox{and} \quad v = (\mu + \epsilon) v_h( \mu + \epsilon) 
\end{equation}
as in Corollary~\ref{cor:harmlrbf} while $C_d$ is the convolution constant for $F_{\mu}$ as
in (\ref{eq:fconv}). 
\end{theorem}

\begin{proof}
With $L \geq 1$ fixed, we will denote by $\tau_t^0 = \tau_t^{h,L}$ and $\tau_t = \tau_t^L$ 
for notational
convenience. Fix $t >0$ and define the function $\Psi_t : [0,t] 
\to \mathcal{B}(\mathcal{H}_{\Lambda_L})$ by setting
\begin{equation} \label{eq:defpsit}
\Psi_t(s) = \left[  \tau_s \left( \tau_{t-s}^0(W(f)) \right), W(g) \right] \, .
\end{equation}
It is clear that $\Psi_t$ interpolates between the commutator associated with the original 
harmonic dynamics, $\tau_t^0$ at $s=0$, 
and that of the perturbed dynamics, $\tau_t$ at $s=t$. A calculation shows that
\begin{equation} \label{eq:dpsit}
\frac{d}{ds} \Psi_t(s) = i \sum_{x \in \Lambda_L} \left[ \, \tau_s \left( \left[ V_x, W ( T_{t-s}f) 
\right] \right) , W(g) \right] \, .
\end{equation}
The inner commutator can be expressed as
\begin{eqnarray}
 \left[ V_x, W ( T_{t-s}f) \right] & = & \int_{\mathbb{C}} \left[ W(z \delta_x), W( T_{t-s}f) \right] 
 \mu_x( dz) \nonumber \\
 & = & W( T_{t-s}f) \mathcal{L}_{t-s;x}(f)  \, .
\end{eqnarray}
where
\begin{equation} \label{eq:defLx}
\mathcal{L}^*_{t-s;x}(f) = \mathcal{L}_{t-s;x}(f) =  \int_{\mathbb{C}} W(z \delta_x) \left\{e^{i {\rm Im} 
\left[ \langle T_{t-s}f, z \delta_x \rangle \right]} -1 \right\} \mu_x(dz) \, \in \mathcal{B}(\mathcal{H}_
{\Lambda_L}) \, .
\end{equation}
Thus $\Psi_t$ satisfies
\begin{equation}
\begin{split}
\frac{d}{ds} \Psi_t(s) =   i \sum_{x \in \Lambda_L} & \Psi_t(s)  \tau_s \left( \mathcal{L}_{t-s;x}(f) 
\right)  \\ 
+ & i \sum_{x \in \Lambda_L} \tau_s \left( W( T_{t-s}f) \right) \, \left[ \tau_s 
\left( \mathcal{L}_{t-s;x}(f) \right), W(g) \right]  \, .
\end{split}
\end{equation}
The first term above is norm preserving. In fact, define a unitary evolution $U_t(\cdot)$ by setting
\begin{equation}
\frac{d}{ds}U_t(s) = - i \sum_{x \in \Lambda_L} \tau_s \left( \mathcal{L}_{t-s;x}(f) \right) U_t(s) 
\quad \mbox{with } U_t(0) = \idty \, .
\end{equation}
It is easy to see that
\begin{equation}
\frac{d}{ds} \left( \Psi_t(s) U_t(s) \right) =  i \sum_{x \in \Lambda_L} \tau_s \left( W( T_{t-s}f) \right) \, 
\left[ \tau_s \left( \mathcal{L}_{t-s;x}(f) \right), W(g) \right] U_t(s) \, ,
\end{equation}
and therefore,
\begin{equation}
\Psi_t(t) U_t(t) = \Psi_t(0) + i \sum_{x \in \Lambda_L} \int_0^t  \tau_s \left( W( T_{t-s}f) \right) \, 
\left[ \tau_s \left( \mathcal{L}_{t-s;x}(f) \right), W(g) \right] U_t(s) \, ds \, .
\end{equation}
Estimating in norm, we find that
\begin{equation} \label{eq:norm1}
\begin{split}
\Big\| \Big[ \tau_t \left(W (f)\right)  , W(g) \Big] \Big\|  \leq &  \Big\| \Big[ \tau_t^0 \left(W (f)\right) , 
W(g) \Big] \Big\| \\
& +  \sum_{x \in \Lambda_L} \int_0^t  \Big\| \left[  \tau_s \left( \mathcal{L}_{t-s;x}(f) \right) , 
W(g) \right] \Big\| \, ds \, .
\end{split}
\end{equation}
Using Corollary~\ref{cor:harmlrbf}, we know that for any $\mu >0$ and $\epsilon >0$, 
\begin{equation} \label{eq:freelrb}
\left\| \left[ \tau_t^0 \left( W(f) \right), W(g) \right] \right\| \, \leq \,
C( \epsilon, \mu) e^{(\mu + \epsilon)v_h( \mu + \epsilon) | t|}  \, \sum_{x, y \in \Lambda_L} \, |f(x)| \, 
|g(y)| \, F_{\mu}(d(x,y)) \, .
\end{equation}
Similarly, one can estimate
\begin{eqnarray}
\left| {\rm Im} \left[  \langle T_{t-s}f, z \delta_x \rangle \right] \right| & \leq 
& |z| \, |T_{t-s}f(x) | \nonumber \\
& \leq & |z| \, C( \epsilon, \mu) e^{(\mu + \epsilon)v_h( \mu + \epsilon) (t-s)}  \, 
\sum_{x' \in \Lambda_L} \, |f(x')| \, F_{\mu}(d(x',x)) \, .
\end{eqnarray}
In this case, the bound
\begin{equation} \label{eq:norm2}
\begin{split}
 \Big\| \left[  \tau_s \left( \mathcal{L}_{t-s;x}(f) \right) , W(g) \right] \Big\| \leq & \, C( \epsilon, \mu) 
e^{(\mu + \epsilon)v_h( \mu + \epsilon) (t-s)}  \, \sum_{x' \in \Lambda_L} \, |f(x')| \, F_{\mu}(d(x',x))  
\times \\
& \quad \times \int_{\mathbb{C}} |z|  \, \Big\| \left[  \tau_s \left( W(z \delta_x) \right) , W(g) \right] 
\Big\| \, |\mu_x|(dz)
\end{split}
\end{equation}
follows from (\ref{eq:defLx}). Setting $c = C( \epsilon, \mu)$ and $v = (\mu + \epsilon)v_h( \mu 
+ \epsilon)$, the combination of (\ref{eq:norm1}), 
 (\ref{eq:freelrb}) , and (\ref{eq:norm2}) demonstrate that
\begin{equation}\label{eq:norm3}
\begin{split}
\Big\| \Big[ \tau_t\left(W (f)\right)  , W(g) \Big] \Big\| \leq \; & c e^{v t} \sum_{x, y 
\in \Lambda_L} |f(x)| \, |g(y)| \, F_{\mu} \left( d(x,y) \right) \\ 
&+ c \sum_{x' \in \Lambda_L} |f(x')| \sum_{x \in \Lambda_L} F_{\mu} \left( d(x,x') \right) \int_0^t  
e^{v (t-s)}  \times \\
&\quad  \times \int_{\mathbb{C}} |z| \, \Big\| \left[  \tau_s \left( W(z \delta_x) \right) , W(g) \right] 
\Big\| \, |\mu_x|(dz) \, ds \,.
\end{split}
\end{equation}
Following an iteration scheme similar to the one in the proof of Theorem~\ref{thm:phi}, one 
arrives at (\ref{eq:anharmbd}) as claimed.
\end{proof}

The statement of the Lieb-Robinson bound proven in Theorem~\ref{thm:ahlrb} can be
strengthened to include a larger class of perturbations. In fact, perturbations involving 
short range interactions can be handled quite similarly. We introduce these perturbations 
as follows.

For each subset $X \subset \Lambda_L$, we associate a finite measure $\mu_X$ on
$\mathbb{C}^X$ and an element $V_X \in \mathcal{B}(\mathcal{H}_{\Lambda_L})$ of the form
\begin{equation} \label{eq:defVX}
V_X = \int_{\mathbb{C}^X} W( z \cdot \delta_X) \, \mu_X(d z) \, ,
\end{equation}
where, for each $z \in \mathbb{C}^X$, the function 
$z \cdot \delta_X : \Lambda_L \to \mathbb{C}$ is given by
\begin{equation}
(z \cdot \delta_X)(x)  = \sum_{x' \in X} z_{x'} \delta_{x'}(x) = \left\{ \begin{array}{cc} z_x 
& \mbox{if } x \in X , \\ 0 & \mbox{otherwise.} \end{array} \right.
\end{equation}
We will again require that $\mu_X$ is invariant with respect to $z \mapsto -z$, and hence, 
$V_X$ is self-adjoint. In analogy to (\ref{eq:defanharmham}), we will write
\begin{equation} \label{eq:defpl2}
V = \sum_{X \subset \Lambda_L} V_X\, ,
\end{equation}
where the sum is over all subsets of $\Lambda_L$.
Here, as before, we will let $\tau^L_t$ denote the dynamics corresponding to $H_L^h +V$. 

The main assumption on these multi-site perturbations is as follows.
We assume there exists a number $\mu_1 >0$ such that for all $0< \mu \leq \mu_1$, 
there is a number $\kappa_\mu >0$ for which given any pair $x, y \in \Lambda_L$, 
\begin{equation} \label{eq:pertbd}
\sum_{\stackrel{X \subset \Lambda_L:}{x,y \in X}} \int_{\mathbb{C}^X} |z_{x}| | z_{y}| \big| 
\mu_X \big|(dz) \leq \kappa_\mu F_\mu \left( d(x,y) \right) \, .
\end{equation} 
In this case, the following Lieb-Robinson bound holds.
\begin{theorem} \label{thm:ahlrbms} Fix $L \geq 1$, $V$, and $\tau_t^L$ as above.
Assume that (\ref{eq:pertbd}) holds. Then, for any $0< \mu \leq \mu_1$ and $\epsilon >0$, 
there exist positive numbers $c$ and $v$ for which the estimate
\begin{equation} \label{eq:anharmbdms}
\left\| \left[ \tau_t^L \left( W(f) \right), W(g) \right] \right\| 
\leq c e^{ (v +  c \kappa_\mu C_d^2 ) |t|} \sum_{x, y \in \Lambda_L} |f(x)| \, |g(y)| 
F_\mu \left( d(x,y) \right)
\end{equation}
holds for all functions $f, g \in \ell^2( \Lambda_L)$ and any $t \in \mathbb{R}$.  
Here $c$, $v$, and $C_d$ are as in Theorem~\ref{thm:ahlrb}.
\end{theorem}
The proof of this result closely follows that of Theorem~\ref{thm:ahlrb}, and so we only 
comment on the differences.
\begin{proof}
For $f,g \in \ell^2(\Lambda_L)$ and $t >0$, define 
$\Psi_t :[0,t] \to \mathcal{B}(\mathcal{H}_{\Lambda_L})$ as in (\ref{eq:defpsit}).
The derivative calculation beginning with (\ref{eq:dpsit}) proceeds as before. Here
\begin{equation} \label{eq:deflzms}
\mathcal{L}_{t-s;X}(f) = \int_{\mathbb{C}^X} W( z \cdot \delta_X) \left\{ e^{i {\rm Im} 
\left[ \langle T_{t-s}f, z \cdot \delta_X \rangle \right] } - 1 \right\} \, \mu_X( d z) \, ,
\end{equation}
is also self-adjoint. The norm estimate
\begin{equation} \label{eq:norm1ms}
\begin{split}
\Big\| \Big[ \tau_t \left(W (f)\right)  , W(g) \Big] \Big\|  \leq  & \Big\| \Big[ \tau_t^0 
\left(W (f)\right) , W(g) \Big] \Big\| \\ & +  \sum_{X \subset \Lambda} \int_0^t 
\Big\| \left[  \tau_s \left( \mathcal{L}_{t-s;X}(f) \right) , W(g) \right] \Big\| \, ds \, ,
\end{split}
\end{equation}
holds similarly. With (\ref{eq:deflzms}), it is easy to see that the integrand in (\ref{eq:norm1ms}) 
is bounded by
\begin{equation} 
c e^{v(t-s)} \sum_{x\in \Lambda_L} \, |f(x)| \, \sum_{x' \in X} F_\mu \left( d(x,x') \right) 
\int_{\mathbb{C}^X}  | z_{x'}|  \,  \, \Big\| \left[  \tau_s \left( W(z \cdot \delta_X) \right) , W(g) \right] 
\Big\| \, |\mu_X|(dz) \, ,
\end{equation}
the analogue of (\ref{eq:norm2}), for any $\mu>0$ and $\epsilon >0$. Proceeding as before, 
\begin{equation}\label{eq:norm4}
\begin{split}
\Big\| \Big[ \tau_t \left(W (f)\right)  , W(g) \Big] \Big\| \leq \; &  c e^{v t} 
\sum_{x, y \in \Lambda_L} |f(x)| \, |g(y)| \, F_\mu \left( d(x,y) \right) \\ &+ c 
\sum_{x \in \Lambda_L} |f(x)| \sum_{X \subset \Lambda_L} \sum_{x' \in X} 
F_\mu \left( d(x,x') \right) \times \\
& \quad \times \int_0^t  e^{v (t-s)}  
\int_{\mathbb{C}^X} |z_{x'}| \, \Big\| \left[  \tau_s \left( W(z \cdot \delta_X) \right) , W(g) \right] 
\Big\| \, |\mu_X|(dz) \, ds \, .
\end{split}
\end{equation}
The estimate claimed in (\ref{eq:anharmbdms}) follows by iteration. In fact, the first term in 
the iteration is bounded by
\begin{equation}
\begin{split}
c \sum_{x\in \Lambda_L} |f(x)| \sum_{X \subset \Lambda_L} \sum_{x_1 \in X} 
& F_\mu \left( d(x,x_1) \right) \int_0^t  e^{v (t-s)}  \\  
& \times \int_{\mathbb{C}^X} |z_{x_1}| \, \Big( c e^{v s} \sum_{x_2 \in X} 
\sum_{y\in \Lambda_L} |z_{x_2}| \, |g(y)| \, F_\mu \left( d(x_2,y) \right) \Big) \, |\mu_X|(dz) \, ds \, \\
\leq  c  t \cdot c e^{v t}  \sum_{x,y \in \Lambda_L} & |f(x)| |g(y)|  
\sum_{x_1, x_2 \in \Lambda_L} F_\mu \left( d(x,x_1) \right)  F_\mu \left( d(x_2,y) \right) \\
& \times \sum_{\stackrel{X \subset \Lambda_L:}{x_1,x_2 \in X}} 
\int_{\mathbb{C}^X} | z_{x_1}| |z_{x_2}| | \mu_{X}|(dz) \, \\
\leq   \kappa_\mu c t  \cdot  c e^{v t} \sum_{x,y \in \Lambda_L} & |f(x)| |g(y)|  
\sum_{x_1, x_2 \in \Lambda_L} F_\mu \left( d(x,x_1) \right) F_{\mu} \left( d(x_1, x_2) \right)  
F_\mu \left( d(x_2,y) \right) \\
\leq  \kappa_\mu C_d^2 c t  \cdot  c e^{v t} \sum_{x,y} & |f(x)| |g(y)| F_\mu \left( d(x,y) \right) \, ,
\end{split}
\end{equation}
where we used that $0< \mu \leq \mu_1$ in the second inequality above. The higher order 
iterates are treated similarly.
\end{proof}

%
%
%
%
%
%
%
%

\section{Existence of the Dynamics}\label{sec:existence}

The goal of this section is to demonstrate that, in a suitable sense, Lieb-Robinson bounds
imply the existence of the dynamics in the thermodynamic limit. We prove a general
statement to this effect in the Section~\ref{sec:bdintex} for the case of bounded interactions.
When considering anharmonic systems, more care must be taken in analyzing the 
thermodynamic limit. We discuss recent results in this direction in Section~\ref{sec:anharmex}.
The analogous problem for the classical anharmonic lattice was analyzed in \cite{lanford:1977}.

\subsection{Bounded Interactions} \label{sec:bdintex}
It is well-known that  Lieb-Robinson bounds for quantum systems
imply the existence of the dynamics in the thermodynamic limit, see e.g. 
\cite{bratteli:1997,nachtergaele:2006}. Here we demonstrate that 
the same basic argument also applies in the general case of bounded interactions. 
Our set-up for this section is similar to that of Section~\ref{subsec:bdint}, 
except that now we regard $\Gamma$, which is still equipped with a metric $d$, 
as a countable set with infinite cardinality.

In many examples, our models are defined over $\Gamma = \mathbb{Z}^d$ for some $d \geq 1$.
For locality estimates and ultimately a proof of the existence of the dynamics,
the underlying lattice structure of $\mathbb{Z}^d$ is not a necessary assumption.
We express the required regularity of $\Gamma$ in terms of a non-increasing
function $F: [0, \infty) \to (0, \infty)$ as mentioned in Section~\ref{subsec:bdint}.

We will say that the set $\Gamma$ is {\it regular} if there exists a non-increasing function
$F:[0, \infty) \to (0, \infty)$ which satisfies:

\noindent i) $F$ is uniformly integrable over $\Gamma$, i.e.,
\begin{equation} \label{eq:fint}
\| \, F \, \| \, := \, \sup_{x \in \Gamma} \sum_{y \in \Gamma}
F(d(x,y)) \, < \, \infty,
\end{equation}

\noindent and

\noindent ii) $F$ satisfies
\begin{equation} \label{eq:intlat}
C \, := \, \sup_{x,y \in \Gamma} \sum_{z \in \Gamma}
\frac{F \left( d(x,z) \right) \, F \left( d(z,y)
\right)}{F \left( d(x,y) \right)} \, < \, \infty.
\end{equation}

For finite sets $X \subset \Gamma$, the Hilbert space $\mathcal{H}_X$ and
the local algebra of observables $\A_X$ are defined as in (\ref{eq:hilX}) and (\ref{eq:algX})
respectively. Recall also that for finite sets $X \subset Y \subset \Gamma$, $\A_X \subset \A_Y$, 
and we may therefore define the algebra of local observables by the inductive limit
\begin{equation}
\mathcal{A}_{\Gamma} \, = \, \bigcup_{X \subset \Gamma} \mathcal{A}_{X},
\end{equation}
where the union is over all finite subsets $X \subset \Gamma$; see
\cite{bratteli:1987,bratteli:1997} for a more detailed discussion.

Our first result on the existence of the dynamics corresponds to Hamiltonians defined as
bounded perturbations of local self-adjoint operators. More specifically, we fix a 
collection of on-site local operators $H^{\rm loc} = \{ H_x \}_{x \in \Gamma}$ 
where each $H_x$ is assumed to be a self-adjoint operator over $\mathcal{H}_x$. 
In addition, we will consider a general class of bounded perturbations.
These perturbations are defined in terms of an interaction $\Phi$, which is a map from the
set of subsets of $\Gamma$ to $\mathcal{A}_{\Gamma}$ with the property that
for each finite set $X \subset \Gamma$, $\Phi(X) \in \mathcal{A}_X$ and
$\Phi(X) ^*= \Phi(X)$. To prove the existence of the dynamics in the
thermodynamics limit, we require a growth condition on the set of interactions $\Phi$ 
being considered. This condition is expressed in terms of a norm analogous 
to the one introduced in our proof of the Lieb-Robinson bounds in Section~\ref{subsec:bdint}. 
Denote by $\mathcal{B}(\Gamma, F)$ the set of interactions
with
\begin{equation} \label{eq:defphia}
\| \Phi \| \, := \, \sup_{x,y \in \Gamma}  \frac{1}{F (d(x,y))} \,
\sum_{X \ni x,y} \| \Phi(X) \| \, < \, \infty.
\end{equation}

Now, for a fixed sequence of local Hamiltonians $H^{\rm loc} = \{H_x \}_{x\in\Gamma}$, as
described above, an interaction $\Phi \in \mathcal{B}(\Gamma, F)$, and a finite subset
$\Lambda \subset \Gamma$, we will consider self-adjoint Hamiltonians of the form
\begin{equation} 
H_{\Lambda} \, = \, H^{\rm loc}_{\Lambda} \, + \, H^{\Phi}_{\Lambda} \, = \, 
\sum_{x \in \Lambda} H_x \, + \, \sum_{X \subset \Lambda} \Phi(X),
\end{equation}
acting on $\mathcal{H}_{\Lambda}$ (with domain given by $\bigotimes_{x \in \Lambda} D(H_x)$
where $D(H_x) \subset \cH_x$ denotes the domain of $H_x$). As these operators are self-adjoint,
they generate a dynamics, or time evolution, $\{ \tau_t^{\Lambda} \}$,
which is the one parameter group of automorphisms defined by
\begin{equation*}
\tau_t^{\Lambda}(A) \, = \, e^{it H_{\Lambda}} \, A \, e^{-itH_{\Lambda}} \quad \mbox{for any} 
\quad A \in \mathcal{A}_{\Lambda}.
\end{equation*}

\begin{theorem}[\cite{nachtergaele:2010}] \label{thm:existbd}
Under the conditions stated above, for all $t \in \bR$ and $A \in \mathcal{A}_{\Gamma}$, 
the norm limit 
\begin{equation}\label{eq:claim} 
\lim_{\Lambda \to \Gamma} \, \tau_t^{\Lambda} (A) = \tau_t(A)
\end{equation} 
exists. Here the limit is taken along any sequence of finite, non-decreasing, 
exhaustive sets $\Lambda$ which tend to $\Gamma$. 
The limiting dynamics $\tau_t$ defines a group of $*$-automorphisms  on the completion of
$\mathcal{A}_\Gamma$. In addition, the convergence is uniform for $t$ in a compact set.
\end{theorem}

Since the convergence proven in Theorem~\ref{thm:existbd} above is in norm and
the estimates provided in Theorem~\ref{thm:phi} are independent of the volume, it is
clear that the dynamics defined above also satisfies the Lieb-Robinson bound
(\ref{eq:lrbd1}). 

\begin{proof} 
Let $\Lambda \subset \Gamma$ be a finite set. Consider the unitary propagator
\begin{equation} \label{eq:intuni}
 \cU_{\Lambda} (t,s) = e^{i t H_{\Lambda}^{\text{loc}} } \, e^{-i (t-s) H_{\Lambda}} \, 
 e^{-is H_{\Lambda}^{\text{loc}}} 
\end{equation}
and its associated {\it interaction-picture} evolution defined by
\begin{equation} \label{eq:intpic}
\tau^{\Lambda}_{t, \text{int}} (A) = \cU_{\Lambda} (0,t) \, A \; \cU_{\Lambda} (t,0) 
\quad \mbox{for all } A \in \mathcal{A}_{\Gamma} \, .
\end{equation} 
Clearly, $\mathcal{U}_{\Lambda}(t,t) = \idty$ for all $t \in \mathbb{R}$, and it is also easy 
to check that 
\begin{equation*}
i \frac{\rd}{\rd t} \, \cU_{\Lambda} (t,s) =  H_{\Lambda}^{\text{int}} (t) \, \cU_{\Lambda} (t,s) 
\quad \mbox{and} \quad
- i \frac{\rd}{\rd s} \, \cU_{\Lambda} (t,s) =  \cU_{\Lambda} (t,s) \, H_{\Lambda}^{\text{int}} (s) 
\end{equation*}
with the time-dependent generator 
\begin{equation} \label{eq:gen}
H^{\text{int}}_{\Lambda} (t) = e^{i H_{\Lambda}^{\text{loc}} t} H_{\Lambda}^{\Phi} 
e^{-i H^{\text{loc}}_{\Lambda} t} = \sum_{Z \subset \Lambda} e^{i H_{\Lambda}^{\text{loc}} t} \, 
\Phi (Z)  \, e^{-i H^{\text{loc}}_{\Lambda} t} \, . 
\end{equation}

Fix $T>0$ and $X \subset \Gamma$ finite. For any $A \in \mathcal{A}_X$, we will show that 
for any non-decreasing, exhausting sequence $\{ \Lambda_n \}$ of $\Gamma$, the sequence
$\{ \tau_{t, \text{int}}^{\Lambda_n}(A) \}$ is Cauchy in norm, uniformly for $t \in [-T,T]$. Since
\begin{equation*}
\tau_t^{\Lambda} (A) = \tau_{t,\text{int}}^{\Lambda} \left(e^{itH_{\Lambda}^{\text{loc}}} \, A \,  
e^{-it H_{\Lambda}^{\text{loc}}} \right) = 
\tau_{t,\text{int}}^{\Lambda} \left(e^{it \sum_{x \in X}H_x} \, A \, e^{-i t \sum_{x \in X} H_x} \right) \, ,
\end{equation*}
an analogous statement then immediately follows for $\{ \tau_t^{\Lambda_n}(A) \}$.

Take $n \leq m$ with $X \subset \Lambda_n \subset \Lambda_m$ and calculate
\begin{equation} \label{eq:diff}
\tau_{t,\text{int}}^{\Lambda_m} (A) - \tau_{t,\text{int}}^{\Lambda_n} (A) = \int_0^t \frac{\rd}{\rd s} 
\left\{ \cU_{\Lambda_m} (0,s) \, \cU_{\Lambda_n} (s,t) \, A \, \cU_{\Lambda_n} (t,s) \, 
\cU_{\Lambda_m} (s,0) \right\} \, ds \, .
\end{equation}
A short calculation shows that
\begin{equation}
\begin{split}
\frac{\rd}{\rd s} \cU_{\Lambda_m} (0,s) & \, \cU_{\Lambda_n} (s,t) \, A \, \cU_{\Lambda_n} (t,s) \, 
\cU_{\Lambda_m} (s,0) \\
& = \, i \mathcal{U}_{\Lambda_m}(0,s) \left[ \left( H^{\text{int}}_{\Lambda_m}(s) - 
H^{\text{int}}_{\Lambda_n}(s) \right), \mathcal{U}_{\Lambda_n}(s,t) \, A \, 
\mathcal{U}_{\Lambda_n}(t,s) \right] \mathcal{U}_{\Lambda_m}(s,0) \\
& = \, i \mathcal{U}_{\Lambda_m}(0,s) e^{is H_{\Lambda_n}^{\text{loc}}} \left[ \tilde{B}(s), 
\tau_{s-t}^{\Lambda_n} \left( \tilde{A}(t) \right) \right] e^{-is H_{\Lambda_n}^{\text{loc}}} 
\mathcal{U}_{\Lambda_m}(s,0) \, ,
\end{split}
\end{equation}
where
\begin{equation} \label{eq:tat}
\tilde{A}(t) = e^{-it H_{\Lambda_n}^{\text{loc}}} A \, e^{it H_{\Lambda_n}^{\text{loc}}} = 
e^{-it H_{X}^{\text{loc}}} A \, e^{it H_{X}^{\text{loc}}} 
\end{equation}
and
\begin{eqnarray} \label{eq:tbs}
\tilde{B}(s) & = & e^{-is H_{\Lambda_n}^{\text{loc}}}\left( H^{\text{int}}_{\Lambda_m}(s) - 
H^{\text{int}}_{\Lambda_n}(s) \right) e^{is H_{\Lambda_n}^{\text{loc}}} \nonumber \\
& = & \sum_{Z \subset \Lambda_m} e^{is H_{\Lambda_m \setminus \Lambda_n}^{\text{loc}}} 
\Phi(Z)  e^{-is H_{\Lambda_m \setminus \Lambda_n}^{\text{loc}}} - \sum_{Z \subset \Lambda_n} 
\Phi(Z) \nonumber \\
& = & \sum_{\stackrel{Z \subset \Lambda_m:}{ Z \cap \Lambda_m \setminus \Lambda_n \neq 
\emptyset}} e^{is H_{\Lambda_m \setminus \Lambda_n}^{\text{loc}}} \Phi(Z)  e^{-is H_{\Lambda_m 
\setminus \Lambda_n}^{\text{loc}}} 
\end{eqnarray}
Combining the results of (\ref{eq:diff}) -(\ref{eq:tbs}), we find that
\begin{equation}
\left\| \tau_{t,\text{int}}^{\Lambda_m} (A) - \tau_{t,\text{int}}^{\Lambda_n} (A)  \right\| \leq \int_0^t 
\left\| \left[ \tau_{s-t}^{\Lambda_n} \left( \tilde{A}(t) \right), \tilde{B}(s)  \right] \right\| \, ds \,
\end{equation}
and by the Lieb-Robinson bound Theorem \ref{thm:phi}, it is clear that
\begin{eqnarray}
&&\left\| \left[ \tau_{s-t}^{\Lambda_n} \left( \tilde{A}(t) \right), \tilde{B}(s)  \right] 
\right\|\\
& \leq &  \sum_{\stackrel{Z \subset \Lambda_m:}{ Z \cap \Lambda_m \setminus 
\Lambda_n \neq \emptyset}} 
\left\| \left[ \tau_{s-t}^{\Lambda_n} \left( \tilde{A}(t) \right),  
e^{is H_{\Lambda_m \setminus \Lambda_n}^{\text{loc}}} \Phi(Z)  
e^{-is H_{\Lambda_m \setminus \Lambda_n}^{\text{loc}}}  \right] \right\| \nonumber \\
& \leq & \frac{2 \| A \|}{C} \left( e^{2 \| \Phi \| C |t-s|} - 1 \right)  \sum_{y \in \Lambda_m 
\setminus \Lambda_n} \sum_{\stackrel{Z \subset \Lambda_m:}{ y \in Z }}  \| \Phi(Z) \| \sum_{x \in X} 
\sum_{z \in Z} F( d(x,z)) \nonumber \\ & \leq &  \frac{2 \| A \|}{C} \left( e^{2 \| \Phi \| C |t-s|} - 1 \right)  
\sum_{y \in \Lambda_m \setminus \Lambda_n} \sum_{z \in \Lambda_m} \sum_{\stackrel{Z \subset 
\Lambda_m:}{ y, z \in Z }}  \| \Phi(Z) \| \sum_{x \in X}  F( d(x,z)) \nonumber \\
& \leq & \frac{2 \| A \| \| \Phi \|}{C} \left( e^{2 \| \Phi \| C |t-s|} - 1 \right)  \sum_{y \in \Lambda_m 
\setminus \Lambda_n}  \sum_{x \in X}  \sum_{z \in \Lambda_m}  F( d(x,z)) F(d(z,y)) \nonumber \\
& \leq & 2 \| A \| \| \Phi \| \left( e^{2 \| \Phi \| C |t-s|} - 1 \right)  \sum_{y \in \Lambda_m \setminus 
\Lambda_n}  \sum_{x \in X} F( d(x,y)) \, .\nonumber
\end{eqnarray}
With the estimate above and the properties of the function $F$, it is clear that
\begin{equation}
\sup_{t \in [-T,T]} \left\| \tau_{t,\text{int}}^{\Lambda_m} (A) - \tau_{t,\text{int}}^{\Lambda_n} (A)  \right\| 
\to 0 \quad \mbox{ as } n, m \to \infty.
\end{equation}
This proves the claim.
\end{proof}

If all local Hamiltonians $H_x$ are bounded, as is the case for quantum spin systems,
the infinite volume dynamics $\{\tau_t\}$, whose existence we proved above, is 
strongly continuous.  If the $H_x$ are allowed to be densely defined unbounded 
self-adjoint operators, 
we only have weak continuity and the dynamics is more naturally defined on a
von Neumann algebra. This can be done when we have a sufficiently nice invariant
state for the model with only the on-site Hamiltonians. Suppose, for example, that
for each $x\in \Gamma$, we have a normalized eigenvector $\phi_x$ of $H_x$.
Then, for all $A\in \mathcal{A}_\Lambda$, for any finite $\Lambda\subset \Gamma$,
define
\begin{equation}
\rho(A)=\langle \bigotimes_{x\in\Lambda}\phi_x, A \bigotimes_{x\in\Lambda}\phi_x\rangle\, .
\end{equation}
$\rho$ can be regarded as a state of the infinite system defined on the norm completion 
of $\mathcal{A}_\Gamma$. The GNS Hilbert space $\mathcal{H}_\rho$
of $\rho$ can be constructed as the closure of 
$\mathcal{A}_\Gamma \bigotimes_{x\in\Gamma}\phi_x$. 
Let $\psi\in \mathcal{A}_\Gamma \bigotimes_{x\in\Gamma}\phi_x$.
Then 
\begin{equation}
\begin{split}
\left\| \left( \tau_t(A) - \tau_{t_0}(A) \right)  \psi \right\| \leq & \left\| 
\left( \tau_t(A) - \tau_t^{(\Lambda_n)}(A) \right)  \psi \right\| \\
+ & \left\| \left( \tau_t^{(\Lambda_n)}(A) - \tau_{t_0}^{(\Lambda_n)}(A) \right)  
\psi \right\| +
\left\| \left( \tau_{t_0}^{(\Lambda_n)}(A) - \tau_{t_0}(A) \right)  \psi \right\| \, ,
\end{split}
\end{equation}
For sufficiently large $\Lambda_n$, the $\lim_{t\to t_0}$ of middle term vanishes 
by Stone's theorem. The two other terms are handled by (\ref{eq:claim}). It is clear
how to extend the continuity to $\psi\in\mathcal{H}_\rho$.

\subsection{Unbounded interactions} \label{sec:anharmex}

In this section, we will prove the existence of the dynamics in the thermodynamic limit
for the bounded perturbations of the harmonic Hamiltonian we considered in 
Section~\ref{subsec:anharmlrbs}.
The existence of this limit was considered in a recent work \cite{amour:2009} where, 
by modifying the
topology, a rigorous analysis of the dynamics corresponding to the anharmonic system in 
the finite volume could be performed in the limit of the volume tending to $\mathbb{Z}^d$.
Here, as in \cite{nachtergaele:2010}, 
we take a different approach. With our method, we regard the finite volume 
anharmonicities as a perturbation of the {\it infinite} volume harmonic dynamics. 
We prove that the limiting anharmonic dynamics retains the same weak continuity as the
infinite volume harmonic dynamics.

\subsubsection{The infinite-volume harmonic dynamics}

It is well-known that the harmonic Hamiltonian defines a
quasi-free dynamics on the Weyl algebra. We briefly review
these notions here and refer the interested reader to \cite{bratteli:1997} 
(see also \cite{nachtergaele:2010}) for more details. 

In general, the Weyl algebra, or CCR algebra, can be defined over any linear
space $\mathcal{D}$ that is equipped with a non-degenerate, symplectic bilinear
form. For the current presentation, it suffices to think of $\mathcal{D}$ as a subspace of 
$\ell^2( \mathbb{Z}^d)$, e.g. $\ell^2( \mathbb{Z}^d)$, $\ell^1( \mathbb{Z}^d)$, or 
$\ell^2( \Lambda)$
for some finite $\Lambda \subset \mathbb{Z}^d$. In this case, the symplectic form
is taken to be ${\rm Im}[ \langle f, g \rangle ]$.

The Weyl operators over $\mathcal{D}$ are defined by associating non-zero elements
$W(f)$ to each $f \in \mathcal{D}$ which satisfy
\begin{equation} \label{eq:invo}
W(f)^* = W(-f) \quad \mbox{for each } f \in \mathcal{D} \, ,
\end{equation}
and
\begin{equation} 
W(f) W(g) = e^{-i {\rm Im}[ \langle f,g \rangle ]/2} W(f+g) \quad \mbox{for all } f, g \in \mathcal{D} \, .
\end{equation}
It is well-known that there is a unique, up to $*$-isomorphism, $C^*$-algebra generated
by these Weyl operators with the property that $W(0) = \idty$, $W(f)$ is unitary
for all $f \in \mathcal{D}$, and $\| W(f) - \idty \| = 2$ for all $ f \in \mathcal{D} \setminus \{0 \}$, 
see e.g. Theorem 5.2.8 \cite{bratteli:1997}. We will denote by 
$\mathcal{W} = \mathcal{W}( \mathcal{D})$ this algebra, commonly known as the 
CCR algebra, or Weyl algebra, over $\mathcal{D}$.

A quasi-free dynamics on $\mathcal{W}(\mathcal{D})$ is a one-parameter
group of *-automorphisms $\tau_t$ of the form
\begin{equation} \label{eq:weylevo}
\tau_t(W(f))=W(T_t f), \quad f\in \mathcal{D}
\end{equation}
where $T_t:\mathcal{D}\to\mathcal{D}$ is a group of real-linear, symplectic
transformations, i.e., 
\begin{equation} \label{eq:sympT}
T_0 = \idty, \quad T_{s+t} = T_s \circ T_t, \quad \mbox{and,} \quad {\rm Im} 
\left[ \langle T_t f, T_t g \rangle \right] = {\rm Im} \left[ \langle f, g \rangle \right] \, .
\end{equation}
Since $\| W(f) - W(g) \| = 2$ whenever $f \neq g \in \mathcal{D}$, such a quasi-free dynamics
will not be strongly continuous; even in the finite volume.

To define the infinite volume harmonic dynamics, we must recall the finite
volume calculations from Section~\ref{sec:weylop}. Let $\gamma : [- \pi, \pi)^d \to \mathbb{R}$
be defined as in (\ref{eq:defgamma}). Take $U$ and $V$ as in (\ref{eq:defU+V}) with 
$\mathcal{F}$ the unitary Fourier transform from $\ell^2(\mathbb{Z}^d)$ to $L^2([-\pi,\pi)^d)$.
Setting
\begin{equation}
T_t = (U+V) \mathcal{F}^{-1} M_t \mathcal{F} (U^*-V^*)
\end{equation}
one can easily verify (\ref{eq:sympT}) using the properties of $U$ and $V$; namely
(\ref{eq:bog1}) and (\ref{eq:bog2}). If, in addition, $\mathcal{D}$ is $T_t$ invariant, 
then Theorem 5.2.8 of \cite{bratteli:1997} guarantees the existence of a unique one 
parameter group of $*$-automorphisms 
on $\mathcal{W}( \mathcal{D})$, which we will denote by $\tau_t$, that satisfies
(\ref{eq:weylevo}). This defines the harmonic dynamics on such a $\mathcal{W}( \mathcal{D})$.

With calculations similar to those found in \cite{nachtergaele:2009a}, one finds that the 
mapping $T_t$ defined above
can be expressed as a convolution, analogously to the finite volume calculations. In fact,
\begin{equation} \label{eq:defft2}
T_tf = f * \left(H_t^{(0)} - \frac{i}{2}(H_t^{(-1)} + H_t^{(1)}) \right) + 
\overline{f}*\left( \frac{i}{2}(H_t^{(1)} - H_t^{(-1)}) \right).
\end{equation}
where
\begin{equation}\label{eq:H2}
\begin{split}
H^{(-1)}_t(x) &= \frac{1}{(2 \pi)^d} {\rm Im} \left[ \int \frac{1}{ \gamma(k)} 
e^{i(k \cdot x-2\gamma(k)t)} \, d k \right],
\\
H^{(0)}_t(x) &= \frac{1}{(2 \pi)^d}  {\rm Re} \left[ \int e^{i(k \cdot x - 2\gamma(k)t)} \, dk \right],
\\
H^{(1)}_t(x) &=  \frac{1}{(2 \pi)^d} {\rm Im} \left[ \int \gamma(k) \, e^{i(k \cdot x-2\gamma(k)t)} \, 
dk \right]  \, ,
\end{split}
\end{equation}
and we have replaced the Riemann sums from the finite volume with integrals.
The following result holds.
\begin{lemma}\label{lem:htx2}
Consider the functions defined in (\ref{eq:H2}). For $\omega\geq 0, \lambda_1,
\ldots,\lambda_d\geq 0$, but such that  
$c_{\omega,\lambda} = (\omega^2 + 4 \sum_{j=1}^d \lambda_j )^{1/2} >0$, 
and any $\mu >0$, the bounds
\begin{equation}
\begin{split}
\left| H_t^{(0)}(x) \right| &\leq  e^{-\mu \left( |x| - c_{\omega,\lambda} \max \left( \frac{2}{\mu} \, , \, 
e^{(\mu/2)+1}\right) |t| \right)}
\\
\left| H_t^{(-1)}(x) \right| &\le  c^{-1}_{\omega,\lambda}e^{-\mu \left( |x| - c_{\omega,\lambda} 
\max \left( \frac{2}{\mu} \, , \, e^{(\mu/2)+1}\right) |t| \right)}
\\
\left| H_t^{(1)}(x) \right| &\le c_{\omega,\lambda}e^{\mu/2}e^{-\mu \left( |x| - c_{\omega,\lambda} 
\max \left( \frac{2}{\mu} \, , \, e^{(\mu/2)+1}\right) |t| \right)}
\end{split}
\end{equation}
hold for all $t \in \mathbb{R}$ and $x \in \mathbb{Z}^d$. Here  $|x| = \sum_{j=1}^{d} |x_i|$.
\end{lemma}

Given the estimates in Lemma~\ref{lem:htx2}, equation (\ref{eq:defft2}), and 
Young's inequality, $T_t$ can be defined as
a transformation of $\ell^p(\mathbb{Z}^d)$, for $p\geq 1$. However,
the symplectic form limits us to consider $\mathcal{D}=\ell^p(\mathbb{Z}^d)$
with $1\leq p\leq 2$.

Mimicking the arguments from the proof of Theorem~\ref{thm:harmlrb}, the above estimates 
yield the following Lieb-Robinson bound for the infinite volume harmonic dynamics $\tau_t$.

\begin{theorem} \label{thm:infharmlrb} For any $\mu >0$ and $\epsilon >0$, there exist 
positive numbers $c$ and $v$
for which the estimate
\begin{equation}\label{eq:lrbinfharm}
\left\| \left[ \tau_t \left( W(f) \right), W(g) \right] \right\| \, \leq \,
c  \, e^{v |t|} \sum_{x, y \in \mathbb{Z}^d} \, |f(x)| \, |g(y)| \, F_{\mu}(|x-y|)
\end{equation}
holds for all functions $f,g \in \ell^2( \mathbb{Z}^d)$ and any $t \in \mathbb{R}$. Here one 
may take
\begin{equation} 
c = \left(1+ c_{\omega,\lambda} e^{(\mu+ \epsilon)/2} + c^{-1}_{\omega,\lambda} \right) \, 
\sup_{s \geq 0} e^{- \epsilon s}(1+s)^{d+1} \, 
\end{equation}
and 
\begin{equation}
v = ( \mu + \epsilon) c_{\omega, \lambda} \max \left( \frac{2}{ \mu + \epsilon}, 
e^{ (\mu + \epsilon) /2 +1} \right) \, 
\end{equation}
with $c_{\omega,\lambda} = (\omega^2 + 4 \sum_{j=1}^{\nu} \lambda_j)^{1/2}$.
\end{theorem}

\subsubsection{Weak continuity and the anharmonic dynamics}

As we indicated in the previous section, the harmonic dynamics is not
strongly continuous, not even when restricted to a finite volume. It is possible, however, 
to show weak continuity of the harmonic dynamics in the GNS-representation
of certain states. If $\rho$ is a regular, $\tau_t$-invariant state on $\mathcal{W}( \mathcal{D})$,
then weak continuity follows from proving continuity of the functions
\begin{equation}
t \mapsto \rho \left( W(g_1) W(T_tf) W(g_2) \right) \quad \mbox{for all } g_1, g_2, f \in \mathcal{D} \, .
\end{equation}
In \cite{nachtergaele:2010}, we verified these properties for the infinite volume ground state 
of the harmonic Hamiltonian, i.e.
the vacuum state for the $b$-operators, defined on $\mathcal{W}( \mathcal{D})$ by setting
\begin{equation}
\rho(W(f)) = e^{ - \frac{1}{4} \| (U^*-V^*)f \|^2} \, , \quad \mbox{for all } f \in \mathcal{D}.
\end{equation}
Alternatively, one could also prove weak continuity of the harmonic dynamics in a 
representation corresponding to equilibrium states at positive temperature.  In either case,
it is precisely such a weakly continuous dynamics to which we add our anharmonic perturbations.

Using Proposition 5.4.1 from \cite{bratteli:1997}, which applies to a general weakly continuous
dynamics, in fact to a $W^*$-dynamical system, we define a perturbed dynamics
as follows. Fix a finite subset $\Lambda \subset \mathbb{Z}^d$. Consider
a perturbation of the form $V_{\Lambda} = \sum_{x \in \Lambda} V_x$ where, for each 
$x \in \Lambda$,
$V_x$ is as defined in (\ref{eq:defvx}) of Section~\ref{subsec:anharmlrbs}. The arguments 
below equally well apply to
the multi-site perturbations, see (\ref{eq:defVX}), considered at the end of 
Section~\ref{subsec:anharmlrbs}, however,
for simplicity, we only state results in the case of on-site perturbations, see 
\cite{nachtergaele:2010} for more details. Proposition 5.4.1
demonstrates that the Dyson series 
\begin{equation} \label{eq:dyson}
\tau_t^{(\Lambda)}(W(f))= \tau_t (W(f))
+ \sum_{n=1}^\infty i^n\!\!\! \int_{0\leq t_1\leq t_2\cdots\leq t}
\!\!\!\!\!\!\!\!\! [\tau_{t_n}(V_\Lambda),[\cdots[\tau_{t_1}(V_\Lambda),\tau_t(W(f))]]] \,  dt_1\cdots dt_n \, 
\end{equation}
is well-defined. Furthermore, $\tau_t^{(\Lambda)}$ is weakly 
continuous, and there is a consistency in the iteratively defined dynamics; 
$\tau_t^{(\Lambda_1 \cup \Lambda_2)}$
can also be constructed by perturbing $\tau_t^{(\Lambda_1)}$ on $\Lambda_2$ given that
 $\Lambda_1 \cap \Lambda_2 = \emptyset$.   
 
 As a consequence of (\ref{eq:dyson}), we prove the following Lieb-Robinson bound in 
 \cite{nachtergaele:2010}.
 
 \begin{theorem} \label{thm:ahlrbinf}
Fix a finite set $\Lambda \subset \mathbb{Z}^d$ and let $\tau_t^{(\Lambda)}$ be as defined 
above. Suppose that
\begin{equation}
\kappa = \sup_{x \in \mathbb{Z}^d} \int_{\mathbb{C}} |z|^2 | \mu_x| (d z) < \infty \, .
\end{equation}
For every $\mu >0$ and $\epsilon >0$, there exist positive numbers $c$ and $v$ for which
the estimate
\begin{equation} 
\left\| \left[ \tau_t^{(\Lambda)} \left( W(f) \right), W(g) \right] \right\| \leq c e^{ (v + c \kappa C_d ) |t|} 
\sum_{x, y \in \mathbb{Z}^d} |f(x)| \, |g(y)| F_{\mu} \left( |x-y| \right)
\end{equation}
holds for all functions $f, g \in \ell^2( \mathbb{Z}^d)$ and any $t \in \mathbb{R}$. 
\end{theorem}

To prove this result, one argues as in the proof of Theorem~\ref{thm:ahlrb} except that
the estimates from Theorem~\ref{thm:infharmlrb} replace those of Corollary~\ref{cor:harmlrbf}. 
The numbers
$c$, $v$, and $C_d$, as well as the function $F_{\mu}$, are exactly as in Theorem~\ref{thm:ahlrb}. 

We can now state our result on the existence of the anharmonic dynamics.
\begin{theorem} Let $\tau_t$ be the harmonic dynamics defined on 
$\mathcal{W}( \ell^1( \mathbb{Z}^d))$.
Take $\{ \Lambda_n \}$ to be any non-decreasing, exhaustive sequence of finite subsets of 
$\mathbb{Z}^d$.
For each $x \in \mathbb{Z}^d$, let
\begin{equation}
V_x = \int_{\mathbb{C}} W(z \delta_x) \mu_x(dz) \, ,
\end{equation}
set $V_{\Lambda_n} = \sum_{x \in \Lambda_n}V_x$, and assume that
\begin{equation}
\sup_{x \in \mathbb{Z}^d} \int_{\mathbb{C}} |z| | \mu_x|(dz) \, < \infty \quad \mbox{and} 
\quad \sup_{x \in \mathbb{Z}^d} \int_{\mathbb{C}} |z|^2 | \mu_x|(dz) \, < \infty \, .
\end{equation}
Then, for each $f \in \ell^1( \mathbb{Z}^d)$ and $t \in \mathbb{R}$, the limit
\begin{equation}
\lim_{n \to \infty} \tau_t^{(\Lambda_n)}(W(f))
\end{equation}
exists in norm. Moreover, the limiting dynamics is weakly continuous.
\end{theorem}

\begin{proof}
To show convergence, we estimate 
$\Vert \tau_t^{\Lambda_n}(W(f)) - \tau_t^{\Lambda_m}(W(f)) \Vert$, 
for $\Lambda_m\subset\Lambda_n$, by considering $\tau_t^{\Lambda_n}$ as a 
perturbation of $\tau_t^{\Lambda_m}$. This gives
\begin{equation}
\tau_t^{\Lambda_n}(W(f)) = \tau_t^{\Lambda_m}(W(f)) + i \int_0^t \tau_s^{\Lambda_n} 
\left( \left[ V_{\Lambda_n \setminus \Lambda_m}, \tau_{t-s}^{\Lambda_m}(W(f)) \right] \right) \, ds \, ,
\end{equation}
and therefore
\begin{equation}
\left\|  \tau_t^{\Lambda_n}(W(f)) - \tau_t^{\Lambda_m}(W(f))  \right\| \leq
 \sum_{x \in \Lambda_n \setminus \Lambda_m} \int_0^{|t|} \left\| \left[ V_x , 
 \tau_{|t|-s}^{\Lambda_m}(W(f)) \right] \right\|  ds \, .
\end{equation}
Using Theorem~\ref{thm:ahlrbinf} we find that 
\begin{eqnarray}
\left\| \left[ V_x , \tau_{|t|-s}^{\Lambda_m}(W(f)) \right] \right\|  & \leq & \int_{\mathbb{C}} \left\| 
\left[ W(z \delta_x) , \tau_{|t|-s}^{\Lambda_m}(W(f)) \right] \right\| | \mu_x| (dz) \nonumber \\
& \leq &  c e^{(v+c \kappa C_d)(|t|-s)} \sum_{y \in \mathbb{Z}^d} |f(y)| F_{\mu}(|y-x|) 
\int_{\mathbb{C}} |z| | \mu_x|(dz) \, .
\end{eqnarray}
Since $f \in \ell^1( \mathbb{Z}^d)$ and $F_{\mu}$ is uniformly integrable, this estimate suffices 
to prove that the
sequence is Cauchy. By observation, the proven convergence is uniform on compact $t$-intervals. 

The claimed weak continuity of the limiting dynamics follows by an $\epsilon/3$ argument
similar to the one provided at the end of Section~\ref{sec:bdintex}.
\end{proof}
%
%
%
%
%
%

\section{The structure of gapped ground states}
\label{sec:gapped_gs}

\subsection{The Exponential Clustering Theorem}
\label{sec:exp_clustering}

The local structure of a relativistic quantum field theory \cite{haag:1996}, 
is provided by the finite speed of light which implies an automatic bound for the 
Lieb-Robinson velocity. This implies decay of correlations in  QFT with a gap 
and a unique vacuum \cite{araki:1962,ruelle:1962,haag:1965}.
Fredenhagen \cite{fredenhagen:1985} proved an exponential bound for this
 decay of the form $\sim e^{-\gamma c^{-1}|x|}$, which corresponds to a 
 correlation length of the form  $\xi\leq c/\gamma$. The gap $\gamma$ is interpreted 
 as the mass of the lightest particle. In condensed matter physics, the same
 relation between the spectral gap and the correlation length is widely
 assumed. The role of the speed of light is played by a propagation speed
 relevant for the system at hand, such as a speed of sound. A strict 
 mathematical relationship, however, only holds in one direction: a unique ground
 state with a spectral gap implies exponential decay of spatial correlations
 under quite general conditions, which in particular imply a finite bound on the
 speed of propagation known as a Lieb-Robinson bound. This was proved
 only relatively recently \cite{nachtergaele:2006a, hastings:2006}, using
 an idea of Hastings \cite{hastings:2004}. 
 
 As a consequence of subsequent improvements of the prefactor of the Lieb-Robinson 
 (see \cite{nachtergaele:2009b}), we now also have better constants in the 
 Exponential Clustering Theorem than in the first results. In particular, for observables
 with large support it is significant that the prefactor is only proportional to the smallest
 of the surface areas of the supports of the two observables. E.g., this is important
 in certain applications (see, e.g., \cite{hastings:2009a, matsui:2010}).

\begin{theorem}[\cite{nachtergaele:2009b}]\label{thm:decay}
Let $\Phi$ be an interaction with $\Vert\Phi\Vert_a<\infty$ for some $a>0$.
Suppose $H$ has a spectral gap $\gamma>0$ above a unique ground state 
$\langle\cdot\rangle$. 

Then, there exists $\mu >0$ and a constant $c=c(F,\gamma)$ such that for all 
$A \in \A_X$, $B \in \A_Y$, 
$$
\left\vert \langle AB\rangle -  \langle A \rangle\, \langle B\rangle\right\vert 
\leq c\Vert A\Vert \, \Vert B\Vert  \min(\partial_\Phi X, \partial_\Phi Y) e^{-\mu d(X,Y)} .
$$
One can take
$$
\mu = \frac{a\gamma}{\gamma+4 \Vert \Phi\Vert_a}\, .
$$
\end{theorem}

Using the Lieb-Robinson bounds for oscillator lattices one can 
also prove an exponential clustering theorem for these systems.

\begin{theorem}[\cite{nachtergaele:2009a}]\label{thm:decay_2}
Let $H$ be the anharmonic lattice Hamiltonian with $\lambda \geq 0$ satisfying
the conditions of Case (ii),  and suppose $H$ has a unique ground state and a spectral gap 
$\gamma>0$ above it.
Denote by $\langle \,\cdot\,\rangle$ the expectation in the ground state.
Then, for all functions
$f$ and $g$ with finite supports $X$ and $Y$ in the lattice, we have the following estimate:
\begin{eqnarray*}
&&\left| \langle W(f) W(g)\rangle - \langle W(f) \rangle \langle W(g)\rangle\right|\\
&&\leq C \Vert f\Vert_\infty \, \Vert g\Vert_\infty\,  \min(|X|,|Y|)
e^{-d(X,Y)/\xi}
\end{eqnarray*}
where $\xi = (4a v+\gamma)/(a\gamma)$ and, if we assume $d(X,Y)\geq \xi$, $C$ is a 
constant depending only on the dimension $\nu$.
\end{theorem}

The central argument in the proof of these theorems is the same. Here, we only provide
a sketch and refer to \cite{hastings:2006,nachtergaele:2006a,nachtergaele:2009b} and 
\cite{nachtergaele:2009a} for the details.

Suppose $H\geq 0$ with unique ground state $\Omega$, $H\Omega =0$, with a gap 
$\gamma>0$
above 0. Let $A\in\A_X$ and $B\in\A_Y$, $d(X,Y)>0$, $a,C,v>0$, such that
$$
\Vert [\tau_t(A),B]\Vert \leq C\Vert A\Vert \,\Vert B\Vert e^{-a(d(X,Y)-v |t|)}\,.
$$
We can assume $\langle \Omega, A\Omega\rangle=\langle \Omega, B\Omega\rangle=0$.
We want to show that there is a $\xi<\infty$, independent of $X,Y, A,B$, s.t.
$$
\vert\langle\Omega, AB\Omega\rangle\vert\leq Ce^{-d(X,Y)/\xi}\,.
$$

For $z\in\Cx$, $\Im z\geq 0$, define
$$
f(z)=\langle\Omega, A\tau_z(B)\Omega\rangle
= \int_\gamma^\infty e^{izE}  d\langle A^*\Omega,P_E B\Omega\rangle\,.
$$
For $T>b>0$, and $\Gamma_T$ the upper semicircle of radius $T$ centered at 0:
$$
f(ib)=\frac{1}{2\pi i}\int_{\Gamma_T} \frac{f(z)}{z-ib} dz\, .
$$
Then
$$
\vert\langle\Omega, AB\Omega\rangle\vert
\leq \limsup_{b\downarrow 0, T\uparrow \infty}
\frac{1}{2\pi }\left\vert \int_{-T}^T \frac{f(t)}{t-ib} dt\right\vert\, .
$$
Next, introduce a Gaussian cut-off and remember $f(t)$:
$$
\vert\langle\Omega,AB\Omega\rangle\vert
\leq \limsup_{b\downarrow 0, T\uparrow \infty}
\frac{1}{2\pi }\left\vert \int_{-T}^Te^{-\alpha t^2} 
\frac{\langle\Omega,A\tau_t(B)\Omega\rangle}{t-ib} 
dt\right\vert + C e^{-\gamma^2/(4\alpha)}
$$
assuming $\gamma>2\alpha b$. For $\alpha(d(X,Y)/v)^2\gg 1$, the Lieb-Robinson
bounds lets us commute $\tau_t(B)$ with $A$ in this estimate. Using the spectral
representation of $\tau_t$, we get
$$
\vert\langle\Omega,AB\Omega\rangle\vert
\leq \limsup_{b\downarrow 0, T\uparrow \infty}
\frac{1}{2\pi }\left\vert \int_\gamma^\infty \int_{-T}^T \!\!\!\!\! dt\frac{e^{-iEt}e^{-\alpha t^2}}{t-ib} 
d\langle B^*\Omega,P_E A\Omega\rangle\right\vert + \mbox{err.}
$$
The $t-$integral can be uniformly bounded by $e^{-\gamma^2/(4\alpha)}$. Optimizing
$\alpha$ gives the bounds stated in Theorems \ref{thm:decay} and \ref{thm:decay_2}.

The condition that the ground state be unique can be relaxed. E.g., one gets the same 
result for each gapped ground states of infinite systems with several disjoint ground states.
One can also derive exponential decay in the average of a set of low-energy states
separate by a gap from the rest of the spectrum, if the number of states in the set does not
grow too fast with increasing system size. Another straightforward extension 
covers models of lattice fermions \cite{hastings:2006}.

The Exponential Clustering Theorem says that a non-vanishing gap $\gamma$ implies a finite
correlation length $\xi$. But one can say more about the structure of the ground state. Motivated
by the goal of devising better algorithms to compute ground states and questions related
to quantum computation, a number of further results have been derived. The best known
is Hastings' proof of the Area Law for the entanglement entropy in one dimension
\cite{hastings:2007}, which used an approximate factorization lemma of the ground state 
density matrices. Before we discuss this result and a generalization of it, we make a 
small detour to Valence Bond Solid (VBS) models and Matrix Product States (MPS).
VBS models were first introduced by Affleck, Kennedy, Lieb, and Tasaki \cite{affleck:1987,
affleck:1988}. MPS are a special case of Finitely Correlated States \cite{fannes:1992}.

The first and best known VBS model is the AKLT model named with the initials
of its inventors. This model, itself motivated by Haldane's work 
\cite{haldane:1983,haldane:1983a},
led to a dramatic change in our outlook on quantum spin chains and the ground states of
quantum spin Hamiltonians in general. Before the AKLT model, practically all our 
understanding of the ground states of quantum spin systems stemmed directly from
exact solutions of special models, primarily Bethe-Ansatz solvable models. The Bethe-Ansatz
solutions are tremendously important in their own right but they had seriously biased
our thinking about more general models. The AKLT model and subsequent generalizations
changed that and led to the much better understanding of generic behaviors of quantum 
spin systems that we now have. So, a small excursion to the AKLT model is certainly
justified.

\subsection{The AKLT model}

The AKLT model is a spin-1 chain with the following isotropic nearest-neighbor 
Hamiltonian:
\be
H^{\mbox{AKLT}}_{[a,b]}= \sum_{x=a}^{b-1}[\frac{1}{3}+\frac{1}{2}{\bf S}_x\cdot {\bf S}_{x+1}
+\frac{1}{6} ({\bf S}_x\cdot {\bf S}_{x+1})^2]
\label{aklt}\ee
acting on $\cH_{[a,b]}=(\Cx^3)^{\otimes(b-a+1)}$, where ${\bf S}_x$ is the vector of 
spin-1 matrices acting on the $x$th factor. A straightforward computation based 
on the representation
theory of SU(2) shows that
$$
\frac{1}{3}+\frac{1}{2}{\bf S}_x\cdot {\bf S}_{x+1}
+\frac{1}{6} ({\bf S}_x\cdot {\bf S}_{x+1})^2 =P^{(2)}_{x,x+1}
$$
where $P^{(2)}_{x,x+1}$ is the orthogonal projection onto 
the spin-2 subspace of two spin 1's at $x$ and $x+1$. Therefore, 
$H^{\mbox{AKLT}}_{[a,b]}\geq 0$.
As we will see in a moment, $\dim\ker H_{[-a,b]}=4$, for all $a<b$. Hence, the ground
state energy of the model vanishes for all finite chains.

The AKLT chain has the three properties that characterize the so-called
Haldane phase:

\begin{itemize}
\item It has a unique ground state for the infinite chain. In particular, for $L\geq 1$,
pick $\psi_L\in\ker H_{[-L,L]}$, with $\Vert \psi_L\Vert =1$. Then, for all finite
$X$ and $A\in\A_X$, one has a limiting expectation value
$$
\omega(A)=\lim_{L\to\infty}\langle \psi_L, A\psi_L\rangle
$$
which is independent of the chosen sequence. It follows that $\omega$ is a translation
and SU(2) invariant state of the quasi-local algebra of observables of the infinite 
chain.
\item  The unique ground state $\omega$ has a finite correlation length: there 
exists $\xi>0, C>0$, s.t., for all $A\in\A_X,B\in\A_Y$
$$
\vert \omega(AB)-\omega(A)\omega(B)\vert\leq C \Vert A\Vert \Vert B\Vert e^{-d(X,Y)/\xi}.
$$
In fact, the bound holds with $e^{-1/\xi}=1/3$ and is optimal.
\item The AKLT chain has a spectral gap above the ground state: there exists $\gamma>0$, 
such that for all $b>a$,  the gap of $H_{[a,b]}$, which equals the smallest strictly positive
eigenvalue $E_1$, satisfies $E_1\geq \gamma$. For the infinite chain this is expressed by
$$
\omega(A^* H_X A)\geq \gamma\omega(A^*A).
$$
for all $X$ and all $A\in\A_X$, with 
$$
H_X=\sum_{\stackrel{\{x,x+1\}}{\{x,x+1\}\cap X\neq \emptyset}} P^{(2)}_{x,x+1}
$$
Using the Density Matrix Renormalization Group (DMRG) \cite{white:1992}, one can compute
$\gamma$ numerically to virtually any desired accuracy. E.g., Huse and White found
$\gamma\sim .4097...$ \cite{white:1993}. It was quickly understood that the DMRG can 
be understood as a variational approximation using Matrix Product States (MPS). Since MPS are 
dense in the set of all states \cite{fannes:1992a}, the error of this approximation can, in 
principle, be made arbitrarily small. See \cite{peschel:1999} for a detailed discussion of the 
DMRG and \cite{schollwock:2005,verstraete:2008} for a recent reviews.
\end{itemize}

Haldane predicted these properties for the integer-spin Heisenberg antiferromagnetic chains.
A proof of the existence of non-vanishing spectral gap, or even of the (slightly) weaker property 
of exponetial decay of correlations in the ground state of the (standard) Heisenberg quantum 
spin chains has so far proved elusive, although some interesting conditional statements
have been obtained \cite{affleck:1986,aizenman:1994}.

The AKLT chain was the first proven example of the existence of the Haldane phase.
This is important, but the impact of the explicit construction of the exact ground state
of the AKLT Hamiltonian has gone a great distance beyond that example. It led to analytic 
and numerical techniques to compute and approximate the  complex entangled states 
that occur in many condensed matter systems (see, e.g., \cite{verstraete:2004,verstraete:2006}).

\subsubsection{The AKLT state and its properties}

Recall the Clebsch-Gordan series for the decomposition of the tensor product of two
irreducible representations of SU(2):
$$ 
D^{(s_1)}\otimes D^{(s_2)}\cong D^{(\vert s_1-s_2\vert)}\oplus D^{(\vert s_1-s_2\vert +1)}
\oplus\cdots\oplus D^{(s_1+s_2)}
$$
Let  $\phi\in\Cx^2\otimes\Cx^2$ be the singlet state given by
$$
\phi=\frac{1}{\sqrt{2}}(\ket{\uparrow\downarrow}-\ket{\downarrow\uparrow}),
$$
and let $W:\Cx^3\to\Cx^2\otimes\Cx^2$ be the isometry implementing the embedding
corresponding to $D^{(1)}\subset D^{(1/2)}\otimes D^{(1/2)}$. For any observable 
of the spin-1 system at a single site, $A\in M_3$, $WAW^*$ is its embedding in 
$M_2\otimes M_2$. Then, for every $n\geq 1$, and any $\ket{\alpha},\ket{\beta}\in\Cx^2$,
define the vector $\psi^{(n)}_{\alpha\beta}\in\cH_{[1,n]}$ by
\be
\psi^{(n)}_{\alpha\beta}=(W^*\otimes\cdots\otimes W^*)
(\ket{\alpha}\otimes\phi\otimes\cdots\otimes \phi\otimes\ket{\beta}).
\label{aklt_vector}\ee
Since $W^*$ intertwines SU(2) representations, so does $W^*\otimes\cdots\otimes W^*$.
In particular, $W^*\otimes\cdots\otimes W^*$ leaves the total spin of any vector
unchanged. Since $\psi^{(n)}_{\alpha\beta}$ is the image of a vector in 
$D^{(1/2)}\otimes D^{(1/2)}$, its total spin does not exceed 1. It follows immediately
that  $\psi^{(n)}_{\alpha\beta}$ is a ground state of $H_{[1,n]}$, because $H_{[1,n]}$
is a sum of projections on the spin-2 states of a pair of neighboring spins:
$$
P_{x,x+1}^{(2)}(W^*\otimes W^*) (\ket{\alpha}\otimes\phi\otimes\ket{\beta})
=0
$$
It is not hard to show that the vectors of the form $\psi^{(n)}_{\alpha\beta}$ in fact span
$\ker H_{[1,n]}$, i.e., all ground states of $H_{[1,n]}$ are of this form.

To show the uniqueness of the thermodynamic limit and the finiteness of the
correlation length, we consider the structure of the expectation of an arbitrary
observable:
$$
\omega_n(A_1\otimes\cdots\otimes A_n)
=\frac{\langle \psi^{(n)}_{\alpha\beta}, A_1\otimes\cdots\otimes A_n
\psi^{(n)}_{\alpha\beta}\rangle}{\langle \psi^{(n)}_{\alpha\beta}, \psi^{(n)}_{\alpha\beta}\rangle}.
$$
Careful inspection reveals that we can write this formula in the following form
\be
\omega_n(A_1\otimes\cdots\otimes A_n)
=C_n\Tr P_\alpha \E_{A_1}\circ \E_{A_2}\circ\cdots E_{A_n}(P_\beta)
\label{fcs}\ee
where, for $A\in M_3$ and $B\in M_2$, $\E_A(B)\in M_2$ is defined as
$$
\E_A(B)= V^* A\otimes BV
$$
with $V:\Cx^2\to\Cx^3\otimes \Cx^2$ defined by
$$
V\ket{\alpha}= c(W^*\otimes \idty_2)(\ket{\alpha}\otimes \phi).
$$
and $c$ and $C_n$ are normalization constants. It follows from the properties
of the singlet vector $\phi$ and the intertwining operator $W^*$, that $V$ is
also an intertwiner. By choosing the constant $c$ we can make $V$ the up to
a phase unique isometry corresponding to the inclusion $D^{(1/2)}
\subset D^{(1)}\otimes D^{(1/2)}$. With this choice it is clear that
$$
\E_{\idty_3}(\idty_2)=\idty_2.
$$
The normalization constant $C_n$ in \eq{fcs} is then simply equal to 1.
One can further check by a simple computation that
$$
\E_{\idty_3}(B)=\frac{1}{2}(\Tr\,B)\idty_2 -\frac{1}{3}(B-\frac{1}{2}\Tr\, B),
$$
which is equivalent to the statement that the linear map $\E_{\idty_3}$
is diagonal in the basis of $M_2$ consisting of $\idty_2$ (with eigenvalue $1$)
and the three spin-1/2 matrices (each with eigenvalue $-1/3$. The $k$th powers
of the $\E_{\idty_3}$ are therefore given by
 \be
(\E_{\idty_3})^k(B)=\frac{1}{2}(\Tr\,B)\idty_2 + \left(-\frac{1}{3}\right)^k(B-\frac{1}{2}\Tr\, B),
\label{transfer_power}\ee
for all $B\in M_2$.
From this property it follows immediately that the thermodynamic limit of
the formula \eq{fcs} exists and is independent of the  choice of $P_\alpha$
and $P_\beta$:
\bea
&&\lim_{n_l\to\infty,n_r\to\infty}
\omega_{n_l+n+n_r}(\underbrace{\idty\otimes\cdots\otimes\idty}_{n_l}\otimes 
A_1\otimes\cdots\otimes A_n\otimes\underbrace{\idty\otimes\cdots\otimes\idty}_{n_r})
\nonumber\\
&&\quad=\frac{1}{2}\Tr\, \E_{A_1}\circ \E_{A_2}\circ\cdots E_{A_n}(\idty_2)
\label{aklt_state}\eea
It is also clear that the convergence is exponentially fast and, by the same 
consideration, that the correlations in this state decay as $(1/3)^{\mbox{distance}}$.

The third essential property of the AKLT model is the non-vanishing spectral
gap. In view of our later discussion of the Area Law and approximate factorization
property of gapped ground states, it useful to present to underlying structure of the
AKLT ground state in a bit more detail.

Let $\omega$ denote the ground state of the infinite AKLT chain defined by
\eq{aklt_state}. Let $\rho_{[a,b]}$ be the density matrix describing the restriction 
of $\omega$ to $\A_{[a,b]}$, i.e., for all $A_a,\ldots,A_b\in M_3$,
$$
\omega_n(A_a\otimes\cdots\otimes A_b)
= \Tr \rho_{[a,b]} A_a\otimes\cdots\otimes A_b .
$$
Then, the rank of $\rho_{[a,b]}$ is 4 (equal to the nullity of $H^{\mbox{AKLT}}_{[a,b]}$).
Let $G_{[a,b]}$ be the orthogonal projection onto the range of $\rho_{[a,b]}$.
Then, again using \eq{transfer_power}, one can show that for $\ell\geq 0,a\geq \ell+1$
\be
\Vert G_{[a-\ell,a+\ell+1]}\left[G_{[1,a]}\otimes G_{[a+1,L]}]-  G_{[1,L]} \right]\Vert
\leq Ce^{-\ell/\xi}
\label{almost_factorizing}\ee
This property allows one to prove a uniform lower bound for the gap 
\cite{fannes:1992,nachtergaele:1996,spitzer:2003}. In brief:
$$
\gamma\geq \frac{1}{2}(1-ce^{-\ell/\xi})\times (\mbox{gap of } H^{\mbox{AKLT}}_{[-\ell,\ell]}).
$$

\subsection{The Area Law for the entanglement entropy}

The AKLT state $\omega$ satisfy an ``area bound'' on the entropy of its local restrictions.
In general, this means that for $X\subset\Lambda$ and $\rho_X\in\A_X$ is the density 
matrix describing the restriction of the state to $\A_X$, then 
\be
S(\rho_X)=-\Tr \rho_x\log\rho_X \leq C \vert \partial X\vert
\label{area_law}\ee
If $X$ is an interval, for the AKLT state we have $S(\rho_X)=\log 4$. This
is a trivial consequence of the fact that 4 is the rank of $\rho_X$.
The Area Law Conjecture: \eq{area_law} holds in general for gapped ground states 
of arbitrary quantum spin systems with bounded spins and bounded finite-range
interactions. The conjecture has been proved by Hastings in the case of one-dimensional
systems \cite{hastings:2007}. The evidence for the higher dimensional case 
comes primarily from the existence of a large class of models (VBS models)
with AKLT-like ground states in arbitrary dimensions and their ground state are 
sometimes called PEPS (Products of Entangled Pairs) \cite{verstraete:2006}.
In one dimension there is a density result stating that VBS states with a similar structure 
as the AKLT state are weakly dense in the set of all pure translation invariant states 
\cite{fannes:1992a}. It is also know that each such state is the unique ground state
of a finite-range Hamiltonian with a non-vanishing spectral gap \cite{fannes:1992}.
If we assume that a similar genericity holds for the higher dimensional VBS states,
the evidence for the Area Law Conjecture is quite strong. The rank of the local
density matrices in a VBS state is bounded by the dimension of the space of 
boundary vectors. In one dimension these are the vectors $\ket{\alpha}$ and $\ket{\beta}$
that appear in \eq{aklt_vector}. This dimension is of the form $d^{\vert\partial X\vert}$,
leading immediately to a bound of the form \eq{area_law}.

The theory of higher-dimensional  VBS models is still in progress. In the next section
we present a result that is consistent with the assumption that the unique 
gapped ground states of finite-range Hamiltonians in higher dimensions may
indeed be well approximated by VBS states. Specifically, we will see that
gapped ground states in general have an approximate product structure
similar to \eq{almost_factorizing}.

\subsection{An approximation theorem for gapped ground states}

We will consider a system of the following type: Let $\Lambda$ be a finite
subset of $\Ir^d$. At each $x\in \Lambda$, we have a finite-dimensional
Hilbert space of dimension $n_x$. Let
$$
H_V=\sum_{\{x,y\}\subset \Lambda,|x-y|=1} \Phi(x,y),
$$ 
with $\Vert \Phi(x,y)\Vert\leq J$. 
Suppose $H_V$ has a unique ground state and denote by $P_0$ the corresponding 
projection, and let $\gamma>0$ be the gap above the ground state energy.

For a set $A \subset \Lambda$, the boundary of $A$, denoted by $\partial A$, is
$$
\partial A = \{ x\in A \mid \mbox{there exists } y\in \Lambda\setminus A,  \mbox{with } |x-y| = 1\}.
$$
and for $\ell\geq 1$ define
$$
B(\ell) = \left\{ x\in \Lambda\mid d(x, \partial A) < \ell \right\}.
$$

\begin{figure}[t]
\begin{tikzpicture}
 \draw[help lines] (0,0) grid (10,8);
 \draw[thick,rounded corners=0.2cm]
    (0.5,0.5)--(5.5,0.5)--(5.5,5.5)--(0.5,5.5)--cycle;
  \draw[dashed,rounded corners=0.2cm]
    (-0.5,-0.5)--(6.5,-.5)--(6.5,6.5)--(-.5,6.5)--cycle;
     \draw[dashed,rounded corners=0.2cm]
    (1.5,1.5)--(4.5,1.5)--(4.5,4.5)--(1.5,4.5)--cycle;
    \draw (5.5,3.7)--(6.5,3.7);
    \draw[below] (6.2,3.7) node(l){$\ell$};
  \draw[below right] (5.3,.8) node(A){$A$};
  \draw[below right] (6.4,.8) node(B){$B$};
    \draw[below right] (10,.8) node(V){$\Lambda$};
\end{tikzpicture}
\caption{The set $A$ and its fattened boundary $B(\ell)$ used in the statement
of Theorem \ref{thm:appgs}.}
\label{fig:}
\end{figure}
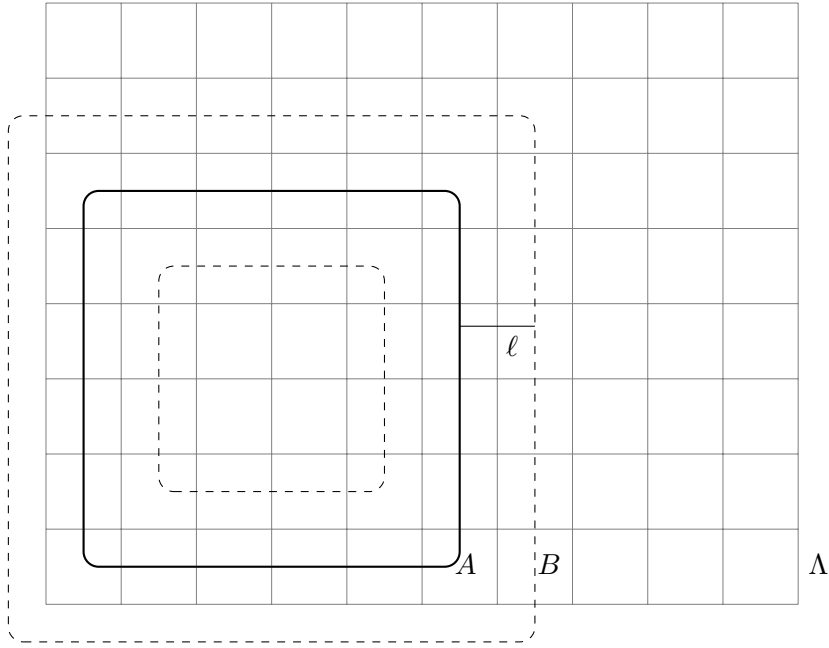

The following generalizes to arbitrary dimensions a one-dimensional result by Hastings 
\cite{hastings:2007}.

\begin{theorem}[\cite{hamza:2009}] \label{thm:appgs}
There exists $\xi>0$ (given explicitly in terms of $d$, $J$, and $\gamma$),
such that for any sufficiently large $m>0$, and any $A\subset \Lambda$,
there exist two orthogonal projections $P_A  \in \A_A$, and
$P_{\Lambda\setminus A} \in \A_{\Lambda\setminus A}$, and  an operator
$P_B\in \A_{B(m)}$ with $\|P_B\| \le 1$, such that
$$
\| P_B (P_A \otimes P_{\Lambda\setminus A}) - P_0 \| \leq C( \xi) | \partial A |^2 e^{-m/ \xi}
$$
where $C(\xi)$ is an explicit polynomial in $\xi$.
\end{theorem}

The proof of this theorem uses several ideas of \cite{hastings:2007}. Below, we only
outline three main steps and refer to \cite{hamza:2009} for the details.

(1) The first step is the bring the Hamiltonian in a form similar to the Hamiltonian
of the AKLT model in the sense that we split the Hamiltonian into terms (three
in this case) which are each individually minimized by the ground state, up to
some error we can make arbitrarily small. The three terms correspond to the
set $A$ and its complement $\Lambda\setminus A$, and a boundary of thickness 
$\ell$ to describe the interaction between $A$ and $\Lambda\setminus A$.
By taking $\ell$ sufficiently large the error can made arbitrarily small.

Without loss of generality we can assume that the ground state energy of $H_\Lambda$
vanishes: $H_\Lambda \psi_0=0$. We aim at a decomposition of $H_\Lambda$, 
for each sufficiently large $\ell$, into three terms:
$$
H_\Lambda= K_A + K_{B(\ell)}+ K_{\Lambda\setminus A}\, ,
$$
with the following two properties for each $K_X$, $X=A,B(\ell),\Lambda\setminus A$:

(i) $\supp K_X\subset X$;

(ii) $\Vert K_X\psi_0\Vert\leq e^{-c\ell}$, for each $X$ and for some $c>0$.

Note that we only assumed $H_\Lambda\psi_0=0$, and no special properties
of the interaction terms $\Phi(x,y)$.
We start from
\be
H_\Lambda= H_I + H_B + H_E\, ,
\label{decomp1}\ee
where 
\beann
I &=& I(\ell) = \left\{ x\in A\mid \mbox{for all }y\in \partial A,
d(x,y) \geq \ell \right\}\\
E &=& E(\ell) = \left\{ x\in V\setminus A\mid \mbox{for all }y\in
\partial A, d(x,y) \geq \ell \right\}.
\eeann
The sets $I(\ell)$ and $E(\ell)$ are the interior and exterior of $A$.
$B(\ell)$ is boundary of thickness $2\ell$:
$$
B(\ell) = \left\{ x\in \Lambda\mid d(x, \partial A) < \ell \right\}.
$$
Note that $\Lambda$ is the disjoint union of $I,B$, and $E$.
Now define
$$
H_I = \sum_{\stackrel{X \subset \Lambda:}{X \cap I \neq \emptyset}}
\Phi(X),
\quad
H_B = \sum_{\stackrel{X \subset \Lambda:}{X \subset B}} \Phi(X),
\quad
H_E = \sum_{\stackrel{X \subset \Lambda:}{X \cap E \neq \emptyset}}
\Phi(X).
$$
For $\ell > 1$, there are no repeated terms and (\ref{decomp1}) holds.
However, there is no guarantee that $\Vert H_X\psi_0\Vert$ will be small.
In general, this will not be the case but we can arrange it so that
each term has 0 expectation in $\psi_0$.  What is needed is a bit of `smoothing'
of the terms using the dynamics: for $X\in\{I,B,E\}$ define
$$
(H_X)_\alpha =  \sqrt{ \frac{ \alpha}{ \pi}} \int_{-
\infty}^{\infty} \tau_t ( H_X ) \, e^{- \alpha t^2} \, dt\, ,
$$
for $\alpha >0$. Since the full Hamiltonian is invariant under the dynamics
it generates, we still have
$$
H_\Lambda= (H_I)_\alpha + (H_B)_\alpha + (H_E)_\alpha\, ,
$$
But now we can show that $\Vert H_X\psi_0\Vert$ is small for $\alpha$ small.
Unfortunately, the support of $(H_X)_\alpha$ is no longer $X$. 
The easiest way to correct this is by redefining them with a suitably restricted 
dynamics as follows:
$$
K_A^{(\alpha)} =  \sqrt{ \frac{ \alpha}{ \pi}} \int_{-
\infty}^{\infty} e^{itH_A} H_Ie^{-itH_A}  \, e^{- \alpha t^2} \, dt\, ,
$$
and similarly define $K_{\Lambda\setminus A}^{(\alpha)}$ using $H_{\Lambda\setminus A}$, 
and $K_B^{(\alpha)} $ using $H_{B(2\ell)}$.

A good choice for $\alpha$ is $av^2/(2\ell)$, where $a$ and $v$ are the constants
appearing in the Lieb-Robinson bounds for the model under consideration. With this
choice one can show that all errors are bounded by
$$
\epsilon(\ell)\equiv C(d,a,v) J^2 \vert\partial A\vert \ell^{d-1/2} e^{-\ell/\xi}
$$
with
$$
\xi=2\max (a^{-1}, av^2/\gamma^2)
$$
To summarize, in step (1) we obtained an approximate decomposition
$$
\Vert H_\Lambda - (K_A^{(\alpha)}+K_B^{(\alpha)}+K_{\Lambda\setminus A}^{(\alpha)})\Vert
\leq \epsilon(\ell)
$$
with the desired property 
$$
\Vert K_{X}^{(\alpha)}\psi_0\Vert \leq \epsilon(\ell)
$$
for $X=A,B,\Lambda\setminus A$.

(2) Next, we define the projections $P_A$ and $P_{\Lambda\setminus A}$
as the spectral projections of  $K_{A}^{(\alpha)}$  and $K_{\Lambda\setminus A}^{(\alpha)}$
projecting onto their eigenvectors  with the eigenvalues $\leq \sqrt{\epsilon(\ell)}$.
This gives
$$
\Vert (\idty - P_A)\psi_0\Vert\leq \frac{1}{\sqrt{\epsilon(\ell)}}\Vert K^{(\alpha)}_A\psi_0\Vert
\leq \sqrt{\epsilon(\ell)}
$$
and similary for $P_{\Lambda\setminus A}$. 
Since the projections commute we have the identity 
$$
2(\idty - P_A P_{\Lambda\setminus A})=(\idty -P_A)(\idty+P_{\Lambda\setminus A})+
(\idty-P_{\Lambda\setminus A} )(\idty +P_A)\, ,
$$ 
from which we obtain
\begin{equation}
\Vert P_0 - P_0 P_A P_{\Lambda\setminus A})\Vert
=\Vert P_0 (\idty - P_A P_{\Lambda\setminus A})\Vert \leq 2\sqrt{\epsilon(\ell)}\, .
\label{result_step_2}\end{equation}

(3) As the final step, we need to replace the ground state projection 
$P_0$ which multiplies  $P_A P_{\Lambda\setminus A}$ in the 
LHS of (\ref{result_step_2}), by an operator supported in the boundary set
$B(m)$ for a suitable $m$.

We start from the observation that for a self-adjoint operator with a gap, such as $H_\Lambda$,
the ground state projection $P_0$ can be approximated by $P_\alpha$ defined by
$$
P_\alpha = \sqrt{\frac{\alpha}{\pi}}\int_{-\infty}^\infty e^{itH_V}e^{-\alpha t^2}dt\, .
$$
If the gap is $\gamma$, and with our choice of $\alpha$, we have
$$
\Vert P_\alpha -P_0\Vert \leq e^{-\gamma^2/(4\alpha)}\leq e^{-\ell/\xi}\, .
$$
We modify this formula for $P_\alpha$ in two ways: 

(i) we replace $e^{itH_\Lambda}$ 
by
$$
e^{it(K_A^{(\alpha)}+K_B^{(\alpha)}+
K^{(\alpha)}_{\Lambda\setminus A})}e^{-it (K_A^{(\alpha)}+
K^{(\alpha)}_{\Lambda\setminus A})}\, ;
$$
This leads to an operator $\tilde P_B$ defined by
$$
\tilde P_B = \sqrt{\frac{\alpha}{\pi}}\int_{-\infty}^\infty
e^{it(K_A^{(\alpha)}+K_B^{(\alpha)}+K^{(\alpha)}_{\Lambda\setminus A})}e^{-it (K_A^{(\alpha)}+
K^{(\alpha)}_{\Lambda\setminus A})}e^{-\alpha t^2}dt
$$

(ii) Then we approximate the result by an operator supported in $B(3\ell)$
to obtain our final results for $P_B$ which appears in the statement
of the theorem:
$$
P_B = \Tr_{\cH_{\Lambda\setminus B(3\ell)}} \tilde P_B\, .
$$
With these definitions it is straightforward to show that both 
$\Vert P_0P_A P_{\Lambda\setminus A} - \tilde P_B P_A P_{\Lambda\setminus A}\Vert$
and $\Vert\tilde P_B -P_B\Vert$  are small. This concludes the outline of the proof
of Theorem \ref{thm:appgs}.

\section{Conclusions and Further Developments}\label{sec:conclusion}

In these lecture notes we have reviewed the derivation of Lieb-Robinson bounds for a
considerable variety of systems, including many of the well-known models frequently
used in condensed matter physics. It is important to continue to expand the class of 
systems for which Lieb-Robinson bounds can be proved. Their relevance keeps growing
as new applications continue to be found.

An application we have not discussed in these notes is the higher-dimensional version
of the Lieb-Schultz-Mattis Theorem. The classical Lieb-Schultz-Mattis Theorem  \cite{lieb:1961}
is for spin-1/2 spin chains and states that if the ground state 
is unique, then the gap above it must vanish at least as $C/L$, where $C$ is a constant
and $L$ is the length of the chain. Later, Affleck and Lieb generalized the result
to other one and quasi-onedimensional models  \cite{affleck:1986}. In particular, their
result applies to those chains of even length with spins having 
arbitrary half-integer magnitude.  But it took more than forty years for someone to
make real progress on a higher-dimensional analogue of the Lieb-Schultz-Mattis
Theorem. In 2004 Hastings found a novel approach using Lieb-Robinson bounds
directly and indirectly (through the Exponential Clustering Theorem), that allowed to
obtain a Lieb-Schultz-Mattis theorem in arbitrary dimension \cite{hastings:2004}.
The result applies to a wide class of Hamiltonians, which includes the half-integer
spin antiferromagnetic Heisenberg model on $\Ir^d$ with suitable boundary conditions
and states that if the ground state is non-degenerate the gap of the system of linear
size $L$, $\gamma_L$, must satisfy:
\begin{equation}
\gamma_L \leq C \frac{\log(L)}{L}.
\end{equation}
The detailed conditions of the theorem and a rigorous proof are given in
\cite{nachtergaele:2007}. An overview can be found in \cite{nachtergaele:2009b}.

A new application of Lieb-Robinson bounds and their
consequences, which recently appeared on the arXiv, is concerned with the Quantum Hall
Effect \cite{hastings:2009a}. This work is concerned with system defined on a two-dimensional
lattice with torus geometry, with interactions that are uniformly bounded and of finite range,
and which preserve charge. The authors of  \cite{hastings:2009a} prove that if such a system 
has a unique ground state with a non-vanishing spectral gap, $\gamma$, above it, its 
Hall conductance, $\sigma_{x,y}$, as defined by the Kubo formula, will show sharp 
quantization. More precisely, for a system of linear size $L$, an estimate of the following 
type is obtained. There is an integer $n$, and constants $C$ and $c>0$, such that
\begin{equation}
\left\vert \sigma_{x,y}-n\frac{e^2}{h}\right\vert\leq C L^{3}e^{-c\gamma^2 L^{2/5}/(\log L)^6}
\end{equation}

An even more recent application is the stability of the Toric Code model 
\cite{kitaev:2003} under small perturbations of the interaction \cite{bravyi:2010}.
This result significantly enhances the plausibility of implementing quantum computation
using topologically ordered ground states. Again Lieb-Robinson bounds and its
corollaries play a crucial role in turning ``adiabatic continuation'', a tool pioneered
by Hastings \cite{hastings:2004,hastings:2005}, into a practical tool for the proof of this result.

\subsection*{Acknowledgments}

The work reported on in this paper was supported by the National Science
Foundation under Grants \#DMS-0757581 (BN) and \#DMS-0757424 (RS).
BN would like to thank the organizers of the Arizona School {\it Entropy and the 
Quantum} for the opportunity to give the lectures on which this paper is based.
The Arizona school was supported by the National Science Foundation under Grant
\#DMS-0852422.

\bigskip

\bibliographystyle{hamsplain}
\providecommand{\bysame}{\leavevmode\hbox to3em{\hrulefill}\thinspace}

\end{document}